\documentclass[a4,12pt]{article}
\usepackage{graphicx}
\usepackage{color}
\usepackage{colordvi}
\usepackage{mathrsfs}
\usepackage{axodraw1}
\advance \headheight by 3.0truept       
\def\sla#1{\rlap/#1}

\newcommand{\lsim}{\raisebox{-0.13cm}{~\shortstack{$<$ \\[-0.07cm] $\sim$}}~}
\newcommand{\gsim}{\raisebox{-0.13cm}{~\shortstack{$>$ \\[-0.07cm] $\sim$}}~}

\begin{document}
\thispagestyle{empty} 
\begin{center}
\vspace{2cm}
{\bf {\Large Potentially Large One-loop Corrections to WIMP Annihilation}
}\\

\vspace*{10mm}

{\bf
M. Drees$^{a}$\footnote{e-mail: drees@th.physik.uni-bonn.de}, 
J.~M.~Kim$^a$\footnote{e-mail: juminkim@th.physik.uni-bonn.de}} and 
{\bf K.~I.~Nagao$^b$\footnote{e-mail: nagao@eken.phys.nagoya-u.ac.jp}}\\[5mm]
{\it
${}^a$Physikalisches Institut and Bethe Center for Theoretical
Physics, Universit\"at Bonn,\\ Nussallee 12, D53115 Bonn, Germany\\ 
\vspace*{2mm}
${}^b$Department of Physics, Nagoya University, Nagoya 464-8602, Japan}

\end{center}

\begin{abstract}

  We compute one--loop corrections to the annihilation of
  non--relativistic particles $\chi$ due to the exchange of a (gauge
  or Higgs) boson $\varphi$ with mass $\mu$ in the initial state. In
  the limit $m_\chi \gg \mu$ this leads to the ``Sommerfeld
  enhancement'' of the annihilation cross section. However, here we
  are interested in the case $\mu \lsim m_\chi$, where the one--loop
  corrections are well--behaved, but can still be sizable. We find
  simple and accurate expressions for annihilation from both $S-$ and
  $P-$wave initial states; they differ from each other if $\mu \neq
  0$. In order to apply our results to the calculation of the relic
  density of Weakly Interacting Massive Particles (WIMPs), we describe
  how to compute the thermal average of the corrected cross
  sections. We apply this formalism to scalar and Dirac fermion
  singlet WIMPs, and show that the corrections are always very small
  in the former case, but can be very large in the latter. Moreover,
  in the context of the Minimal Supersymmetric Standard Model, these
  corrections can decrease the relic density of neutralinos by more
  than 1\%, if the lightest neutralino is a strongly mixed state.

\end{abstract}
\clearpage
\setcounter{page}{1}

\section{Introduction}

The existence of non--baryonic Dark Matter (DM) in the universe is by
now well established \cite{Silk_rev}. Recent observations determine
the universal average DM density quite accurately. The exact value and
its uncertainty depend somewhat on the assumptions made in the fit
(``priors''). For example, within a minimal $\Lambda$CDM model, and
combining data on the cosmic microwave background (CMB) anisotropies
with observations of supernovae of type 1a and with analyses of baryon
acoustic oscillations, one finds \cite{Komatsu2009}:
\begin{equation} \label{range}
\Omega_{\rm CDM}h^2 = 0.1131\pm 0.0034 \,.
\end{equation}
Here $\Omega_{\rm CDM}$ is the energy density of cold dark matter in
units of the critical density, and $h$ is the scaled Hubble parameter
such that $H_0=100 \, h \,\mathrm{km}\, \mathrm{sec^{-1}}\,
\mathrm{Mpc^{-1}}$ where $H_0$ is the current Hubble
parameter. Introducing additional parameters in the fit can increase
the allowed range; for example, significantly larger values of
$\Omega_{\rm CDM} h^2$ are allowed if the number of particles that
were relativistic when the CMB decoupled is kept free
\cite{Komatsu2009}. However, the minimal model describes the data
well. Moreover, quite soon data from the Planck satellite are expected
to reduce the error on $\Omega_{\rm CDM} h^2$ to the level of 1.5\%
using CMB measurements alone \cite{planck}.

From the particle physics point of view, Weakly Interacting Massive
Particle (WIMPs) are among the most attractive DM candidates. In
standard cosmology their thermal relic density is naturally of the
right order of magnitude. Owing to their weak interactions, they can
be probed through both direct and indirect detection experiments
\cite{Silk_rev}, although no convincing signal has yet been
found. Moreover, the existence of WIMPs is {\em independently}
motivated in many extensions of the Standard Model (SM) of particle
physics, the most prominent example being extensions based on
supersymmetry (SUSY).

In order to fully exploit the precision of present cosmological data,
the theoretical error on the prediction for the WIMP relic density
should not exceed the error inferred from observations. For a given
cosmological model, the WIMP relic density is determined uniquely by
its annihilation cross section into SM particles, if all WIMPs were
produced thermally. This requires that the post--inflationary Universe
was sufficiently hot, with temperature exceeding about 5\% of the WIMP
mass. Moreover, one has to assume that no other, heavier particles
decayed into WIMPs after WIMP decoupling. Under these conditions,
which are satisfied for standard cosmology, the uncertainty of the
current WIMP relic density is essentially given by the uncertainty of
the WIMP annihilation cross section. In order to calculate this cross
section to percent level accuracy, at least leading radiative
corrections will have to be included.\footnote{These corrections also
  contribute to the cross sections for WIMP annihilation into SM
  particles in our galaxy at present times, which affect the size of
  indirect WIMP detection signals. However, currently there are very
  large uncertainties in the backgrounds to these signals; in case of
  charged particles, propagation through the galaxy adds additional
  uncertainty. Percent level correction to signals for indirect WIMP
  detection are therefore not significant.}

In this paper we calculate one class of potentially sizable radiative
corrections, which are due to the exchange of a boson between the
WIMPs prior to their annihilation. If the mass $\mu$ of the exchanged
boson vanishes, the one--loop expression diverges in the limit of
vanishing relative velocity $v$ between the annihilating WIMPs. These
large ``Sommerfeld'' corrections then have to be re--summed to all
orders in the relevant coupling \cite{summing, Iengo2009, Cassel2009}
(for earlier, related work see \cite{hisano}). However, since WIMPs
couple neither to photons nor to gluons, a new light boson with
sizable coupling to the WIMPs, but not to SM particles, has to be
introduced. The required hierarchy $\mu \ll m_\chi$, where $m_\chi$ is
the mass of the WIMP, then raises naturalness issues.

Here we instead study the case $\mu \lsim m_\chi$. Examples are the
exchange of the light Higgs boson in the minimal supersymmetric
extension of the SM (MSSM), and the exchange of the $Z$ boson in most
``generic'' WIMP models. If $\mu \lsim m_\chi$ and $v \ll 1$ these
corrections can be significantly bigger than the ``generic'' estimate
$\alpha/\pi \sim$ (a few) $\times 10^{-3}$, while remaining small
enough to permit simple one--loop calculations (and without
threatening unitarity \cite{ralston}). We calculate these corrections
in the non--relativistic limit, following Iengo
\cite{Iengo2009}. Since WIMPs decouple at temperature $T \simeq m_\chi
/ 20$, an expansion in the relative velocity usually (but not always
\cite{Griest1991}) converges quite fast, and can therefore be used in
the calculation of the one--loop correction. We also show how to
compute the relevant thermal average over the corrected cross
sections. We first apply our formalism to simple models with scalar or
fermionic SM singlet WIMPs; the corrections are always small in the
former case, but can be sizable in the latter scenario. We then
analyze neutralino annihilation in the MSSM, and show that these
corrections can reduce the relic density of strongly mixed neutralinos
by more than 1\%, comparable to the projected uncertainty of the value
to be inferred from Planck data.

The rest of this paper is organized as follows. In
Sec.~\ref{sec:corr_A} we introduce the formalism, which was initially
proposed to treat non--perturbative Sommerfeld enhancement
\cite{Iengo2009}. In particular, we factorize the correction to the
amplitude due to the boson exchange between two incoming particles. In
Sec.~\ref{sec:relicdensity} we briefly review the standard calculation
of dark matter relic density, and describe the thermal averaging of
the corrected cross section. Numerical results for three WIMP models
will be presented in Sec.~\ref{sec:numerical_calculation}. Finally we
will summarize. Some details of our numerical procedure are described
in Appendix A, while Appendix B shows that our approximate treatment
indeed reproduces the leading terms of an exact calculation of
radiative corrections associated with the initial state in a simple
scalar model.

\section{Correction to the annihilation amplitudes}
\label{sec:corr_A}

\begin{figure}[h!]
\hfill
\begin{minipage}{6.6cm}
\includegraphics[keepaspectratio=true,width=65mm]{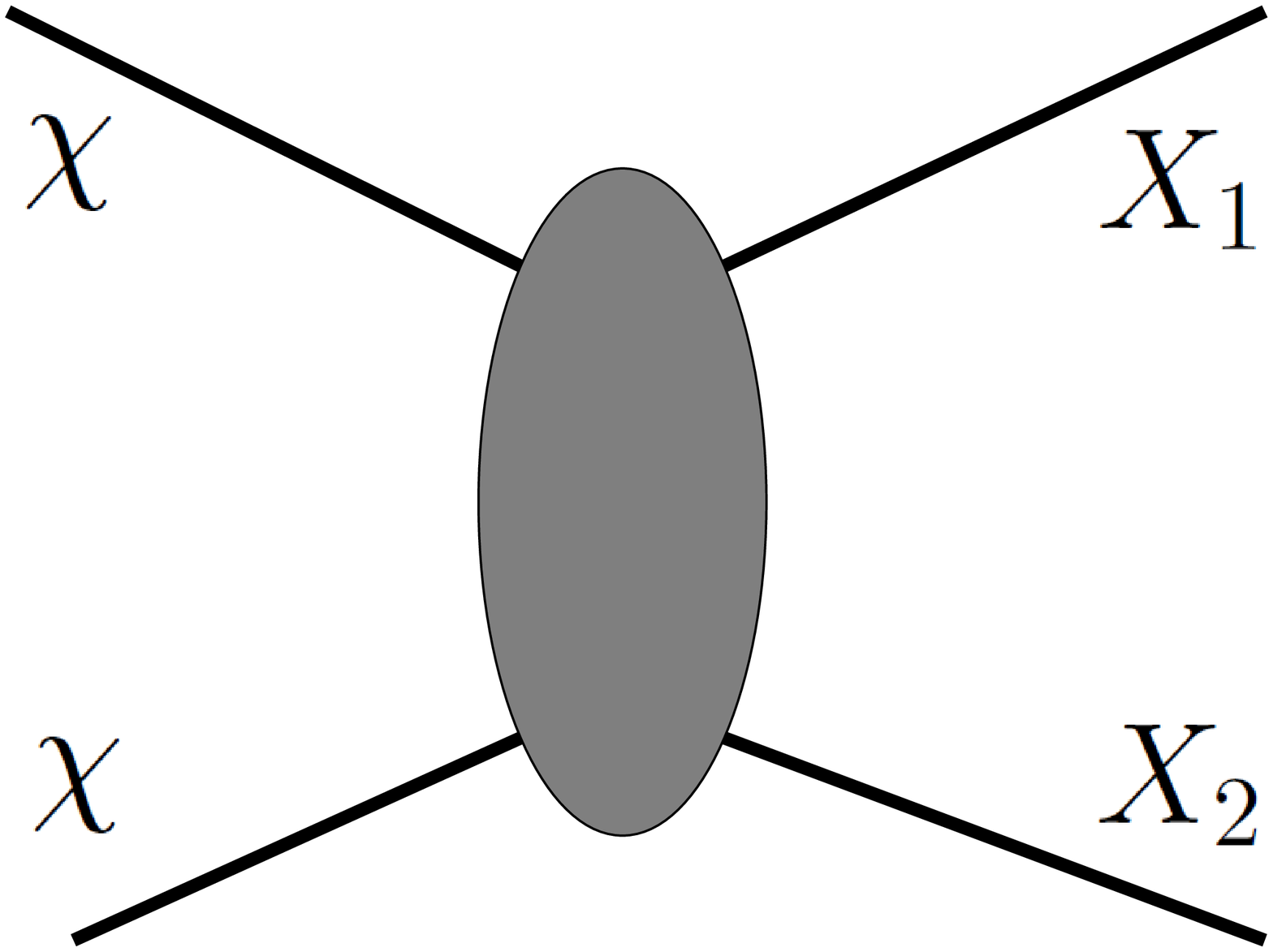}
\caption{Tree--level WIMP annihilation. The grey blob represents
  the exchange of some particle in the $s-,\, t-$ or $u-$channel.} 
\label{fig:a}
\end{minipage}
\hfill
\begin{minipage}{6.6cm}
\includegraphics[keepaspectratio=true,width=65mm]{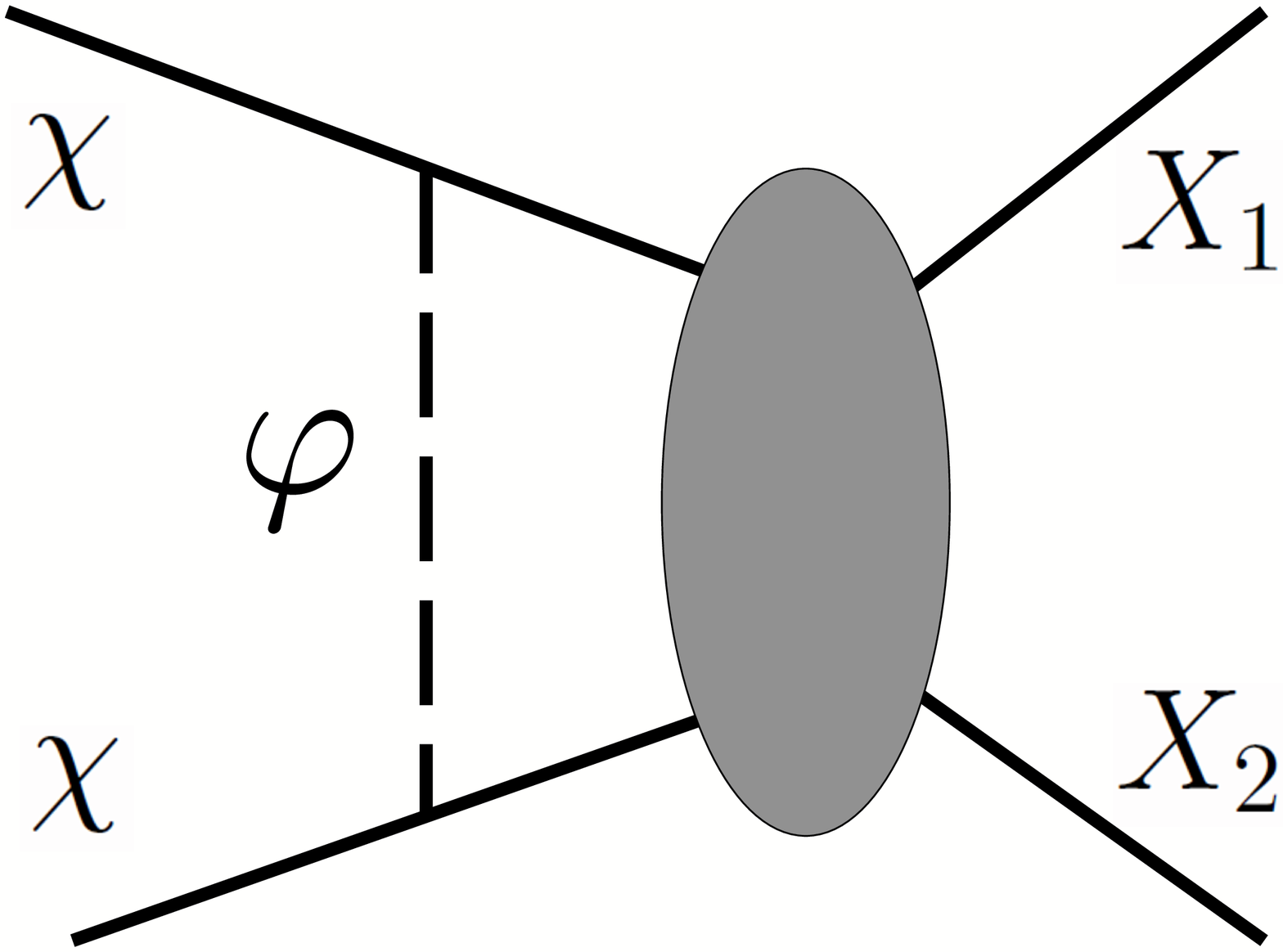}
\caption{The one--loop corrections we are considering in this article.
  The meaning of the grey blob is the same as in Fig.~1.} 
\label{fig:b}
\end{minipage}
\hfill
\end{figure}

Consider the annihilation of two WIMPs $\chi$ into two SM particles:
\begin{equation}
\chi(p_1)+\chi(p_2)\to X_1(p'_1)+X_2(p'_2)\,.
\end{equation}
Generic tree--level diagrams contributing to this process have the
form shown in Fig.~\ref{fig:a}. We want to compute one--loop
corrections of the kind shown in Fig.~\ref{fig:b}, where a boson
$\varphi$ is exchanged between the WIMPs before they annihilate, by
adapting the formalism of Iengo \cite{Iengo2009,itz}. We assume that
$\chi$ is a Majorana fermion; however, in the non--relativistic limit
this will be relevant only for the case where the exchanged boson has
axial vector couplings (see below).

In this formalism $\varphi$ exchange and $\chi$ annihilation are
factorized; the former can then also be understood as re--scattering
of the incoming WIMPs prior to their annihilation. This factorization
($\varphi$ exchange before $\chi$ annihilation) is only expected to
work in the non--relativistic limit. Moreover, it requires the
virtuality of the $\varphi$ propagator to be (much) smaller than that
of the particle exchanged in $\chi$ annihilation. The latter can
always be satisfied for $\mu \ll m_\chi$, where $\mu$ is the mass of
the exchanged boson, but it can also be satisfied for $\mu \sim
m_{\chi}$ if the WIMPs annihilate through the exchange of a particle
$Y$ with mass $M_Y \gg m_\chi$. However, we will see that the
corrections become small if $\mu \gsim m_\chi$. Note also that these
corrections do {\em not} capture UV effects like the renormalization
of the tree--level couplings of $\chi$.

Let $P = (p_1+p_2)/2 = (p'_1+p'_2)/2$, $p=(p_1-p_2)/2$ and $p' =
(p'_1-p'_2)/2$; recall that $p_1,\,p_2$ and $p'_1,\,p'_2$ are the
four--momenta in the initial and final state, respectively. In the cms
we have $P_0=\sqrt{\vec{p}^2+m_\chi^2}, \vec{P} = 0$ and $p_0 =
0$. The initial state kinematics is thus fixed by $\vec{p}$. We write
the one--loop corrected amplitude for annihilation from the partial
wave denoted by $L$ as
\begin{equation}
A_L(|\vec{p}|, p') = A_{0,L}(|\vec{p}|, p') + \delta A_L(|\vec{p}|, p')\,,
\label{eq:corr_A}
\end{equation}
where $A_{0,L}$ and $\delta A_L$ denote the tree--level amplitude and
the one--loop correction term, respectively. We are only interested in
the cases $L=S$ and $L=P$. The contribution of higher partial waves is
strongly suppressed during WIMP freeze--out. 

The corrections can be calculated starting from the observation that
annihilation from an $S- \ (P-)$wave initial state can be described by
a pseudoscalar (scalar) $\chi$ current \cite{dn}. This allows to write
the correction as a standard vertex (three--point function)
correction. Let us begin with the simple case of scalar boson exchange:
\begin{eqnarray} \label{en1}
\delta A_L(|\vec{p}|, p') &=& i g^2 \bar{v}(p_2) \int \frac {d^4
q}{(2\pi)^4} \frac { \sla{q} - \sla{P} + m_\chi } { (q-P)^2 - m_\chi^2
+ i \epsilon } \left(\gamma_5\right)^{n_L} \frac { \sla{q} + \sla{P} +
m_\chi } { (q+P)^2 - m_\chi^2 + i \epsilon } 
\nonumber \\ & & \hspace*{25mm} \times
\frac {1} {(p-q)^2 - \mu^2 + i \epsilon } \tilde{A}_{0,L}(|\vec{q}|,
p') \, u(p_1)\,. 
\end{eqnarray}
Here $g$ is the strength of the coupling between the boson and the
WIMP, $n_L = 1 \ (0)$ for annihilation from an $S- \ (P-)$wave initial
state, and the reduced tree-level amplitude $\tilde A_{0,L}$ describes
the blob in Fig.~1 [except for the factor $\gamma_5$ in case of
$S-$wave annihilation, which appears explicitly in Eq.(\ref{en1})] as well as
the final state.\footnote{Note that Eq.(\ref{en1}) is one--loop exact,
  if the $\vec{q}$ dependence of $\tilde A_{0,L}$ is kept. For a full
  non--perturbative treatment the complete reduced amplitude $\tilde
  A_L$ should appear again in the integral on the right--hand side
  \cite{Iengo2009}. Recall, however, that we are only interested in
  one--loop corrections, in which case we may use $\tilde A_{0,L}$ in
  the integrand.} We will later show how to treat the exchange of
spin--1 bosons.

As a first simplification, one then uses the fact that the full
relativistic boson propagator satisfies $1/(k^2 - \mu^2) =
-1/(\vec{k}^2 + \mu^2) \left[ 1 - k_0^2/ (k^2 - \mu^2) \right]$; the
second term can be omitted in the non--relativistic limit, where the
energy exchange is much smaller than the three--momentum
exchange. Moreover, to leading order in the non--relativistic
expansion, the $\vec{q}-$dependence of the reduced amplitude can be
neglected, i.e. the factor $\tilde{A}_{0,L}$ can be pulled out of the
integral.\footnote{In the $P-$wave case, the nontrivial dependence on
  the initial state three--momentum stems from the spinors describing the
  initial state and the Dirac structures shown explicitly in
  Eq.(\ref{en1}), not from the reduced amplitude. $\tilde{A}_{0,L}$ in
  fact does not depend on $|\vec{q}|$ if $\chi \chi$ annihilation
  proceeds through an $s-$channel diagram. For $t-$ or $u-$channel
  annihilation we have to assume that the particle exchanged in the
  annihilation process is significantly more off--shell than
  $|\vec{q}| \sim |\vec{p}|$.} We then perform the
integrals in Eq.(\ref{en1}) starting from the $q_0$ integration. We
are looking for poles in the lower half--plane, with residues that
diverge in the limit $\vec{q}, \vec{p} \rightarrow 0$
\cite{Iengo2009}.  This gives
\begin{equation} \label{en2}
\delta A_L(|\vec{p}|, p') \simeq  g^2 \bar{v}(p_2) \int \frac {d^3
q}{(2\pi)^3} \frac { \left( \sla{q} - \sla{P} + m_\chi \right) 
\left(\gamma_5\right)^{n_L} \left( \sla{q} + \sla{P} + m_\chi \right) } 
{8 \omega P_0 \left( \omega - P_0 \right) \left[ \left( \vec{p} -
      \vec{q} \right)^2 + \mu^2 \right] } \tilde{A}_{0,L}(\vec{p}')
\, u(p_1)\,,
\end{equation}
where $\omega = \sqrt{\vec{q}^2 + m_\chi^2}$; in the numerator one
should take $q_0 = \omega - P_0$. For small 3--momenta, we have
\begin{equation} \label{en3}
\omega - P_0 \simeq \frac {\vec{q}^2 - \vec{p}^2} {2 m_\chi}\,;
\end{equation}
note that this vanishes for $\vec{p}, \vec{q} \rightarrow 0$, as
advertised. 

To zeroth order in the non--relativistic expansion we can set $\vec{q}
\rightarrow 0$ in the numerator of Eq.(\ref{en2}). We see that this
gives a non--vanishing result only if $n_L = 1$, i.e. if a $\gamma_5$
matrix is present; recall that this corresponds to annihilation from
an $S-$wave. To this order we can replace $\sla{P} = P_0 \gamma^0$
acting to the right (on the $u-$spinor) by $m_\chi$. The numerator of
Eq.(\ref{en2}) then reduces to $4 \gamma_5 m_\chi^2$. Note that the
factor of $\gamma_5$ is required in order to be able to access the
large component of $\bar{v}(p_2)$; this is most easily seen in the
Dirac representation. Moreover, the factor $\omega P_0$ in the
denominator of Eq.(\ref{en2}) can be replaced by $m_\chi^2$, up to
corrections which are of second order in three--momenta.

To summarize, we have made three approximations:
\begin{enumerate}
\item We ignored the energy dependence of the $\varphi$ propagator.
\item In the $q_0$ integral we only kept the pole with the leading
  residue (in the non--relativistic limit).
\item We ignore all $\vec{q}$ dependence in the numerator (for
  annihilation from the $S-$wave).
\end{enumerate}
Note that these approximations have to be taken simultaneously in
order to get a UV--finite result. One may worry that this gives a
rather poor approximation of the exact vertex correction even in cases
where the latter are finite, as in a purely scalar theory. In Appendix
B we show that this approximation can indeed differ substantially from
the exact vertex correction if $\mu \ll m_\chi$.  Note that the exact
vertex correction by itself becomes IR--divergent for $\mu = 0$. This
leads to terms $\propto \log (m_\chi/\mu)$ which appear in the exact
vertex correction, but not in our approximation. However, these terms
are canceled by real emission diagrams and wave function corrections,
which have to be included in order to obtain an IR--finite result for
$\mu = 0$. In Appendix B we show that, at least for a simple scalar
model, our approximation {\em does} accurately reproduce the {\em
  exact} radiative correction associated with the initial state,
whenever these corrections are large.

Let us therefore proceed with our calculation, which does not require
additional approximations. The angular integrations are
straightforward. One is then left with a single integral to describe
the correction to $S-$wave annihilation:
\begin{equation} \label{eq:deltaAS}
\delta A_S(|\vec{p}|,p') = \frac {g^2} {(2\pi)^3}
 \frac{\pi m_\chi} {|\vec{p}|} A_{0,S} 
\int^\infty_0 d|\vec{q}| \frac{ |\vec{q}| } {\vec{q}^2-\vec{p}^2} 
\ln{ \frac {(|\vec{p}| + |\vec{q}| )^2 + \mu^2} 
{(|\vec{p}|- |\vec{q}|)^2 + \mu^2} } \,.
\end{equation}
Note that we have absorbed the spinors into the full tree--level
amplitude $A_{0,S}$. The one--loop correction to the cross section
emerges from the interference between the correction $\delta A_S$ and
the tree--level term $A_{0,S}$. We can thus write
\begin{equation} \label{eq:A_S}
\left. \delta A_S(|\vec{p}|,p') \right|_{\rm 1-loop} = \frac {g^2}
{4\pi^2} \frac{1} {v} A_0(|\vec{p}|,p') I_S(r)\,,
\end{equation}
where $v$ is the relative velocity between the two annihilating WIMPs
in their center of mass frame (i.e., $|\vec{p}| = m_\chi v / 2$), and
we have defined the function
\begin{equation} \label{eq:I_S}
I_S(r) = \Re{\rm e} \left[ \int_0^\infty dx \frac {x} {x^2-1}
\ln{\frac{(1+x)^2+r} {(1-x)^2+r}} \right]\, .
\end{equation}
Here $x=|\vec{q}|/|\vec{p}|$ and $r=\mu^2/\vec{p}^2$.

So far we have assumed that a spin--0 boson with scalar coupling is
exchanged between the annihilating WIMPs. In order to treat more
general cases, we rewrite the numerator of Eq.(\ref{en2}) as
\begin{equation} \label{en4}
{\cal N} = \Gamma \left( \sla{q} - \sla{P} + m_\chi \right) 
\left(\gamma_5\right)^{n_L} \left( \sla{q} + \sla{P} + m_\chi \right)
\bar{\Gamma}\,,
\end{equation}
where $\Gamma$ describes the Dirac structure of the $\varphi \chi
\chi$ coupling and $\bar{\Gamma} = \gamma^0 \Gamma^\dagger \gamma^0$
its Dirac conjugate. Scalar exchange corresponds to $\Gamma = \bar
\Gamma = 1$.

It is easy to see that pseudoscalar exchange, $\Gamma = - \bar
\Gamma = \gamma_5$, does not lead to enhanced contributions. For
example, for $n_L = 1$ and $q \rightarrow 0$ one finds a result
$\propto \gamma_5 \left( m_\chi - \sla{P} \right)^2$, which is ${\cal
  O}(\vec p^2)$. 

Vector exchange corresponds to $\Gamma = \gamma^\nu, \ \bar \Gamma =
\gamma_\nu$, where the Lorentz index $\nu$ has to be summed. For $n_L
= 1$ and $q \rightarrow 0$ this gives 
$${\cal N}_{\rm vector} = - 2 m_\chi \gamma_5 \gamma^\nu \left( m_\chi
  + \sla{P} \right) \gamma_\nu = - 4 m_\chi \gamma_5 \left( 2 m_\chi -
  \sla{P} \right)\,. $$
Again replacing $\sla{P}$ by $m_\chi$ this leads to the {\em same}
result as for scalar exchange, except for an overall sign. However,
this sign is compensated by the extra minus sign in the vector boson
propagator. We thus reproduce the well--known result that in the
non--relativistic limit, the exchange of a vector boson has the same
effect as that of a scalar boson.

Finally, axial vector exchange is described by setting $\Gamma =
\gamma^\nu \gamma_5, \ \bar \Gamma = \gamma_\nu \gamma_5$, where
summation over $\nu$ is again implied. In the $S-$wave case, where
$n_L = 1$ and $q \rightarrow 0$ in the numerator, this gives
$${\cal N}_{\rm axial \ vector} = \gamma_5 \gamma^\nu \left( \sla{P} -
  m_\chi \right)^2 \gamma_\nu = 4 m_\chi \gamma_5 \left( 2 m_\chi +
  \sla{P} \right)\,.$$ 
Again replacing $\sla{P}$ by $m_\chi$, and accounting for the minus
sign in the spin--1 propagator, we find that axial vector exchange
differs from scalar or vector exchange by a factor of $-3$.

This does not seem to have been noticed in the recent literature. We
therefore checked it in the limit where $\varphi$ exchange can be
treated as re--scattering of the incoming WIMPs, leading again to two
on--shell WIMPs. To leading order in velocity expansion we are
interested in the limit of vanishing momentum exchange. However, we
have to keep in mind that the two WIMPs will have to annihilate
through a $\gamma_5$ vertex (for the $S-$wave case). This requires
that the non--relativistic $u$ and $v$ spinors have the same spin. The
rescattering process is then described by the quantity (we omit the
bosonic propagator and all couplings)
\begin{equation} \label{en5}
A_{\rm res} = \sum_{s'} \bar v(p_2,s) \bar{\Gamma} v(p_2,s') \bar
u(p_1,s') \Gamma u(p_1,s)\,,
\end{equation}
where $s,s'$ describe the spin.  Scalar boson exchange again
corresponds to $\Gamma = \bar \Gamma = 1$. In this case only $s' = s$
contributes in the non--relativistic limit, where (in the cms) $p_1 =
p_2 \simeq (m_\chi,\vec{0})$, and one has $A_{\rm res,\ scalar} = -4
m_\chi^2$. One immediately sees that pseudoscalar exchange only
contributes at ${\cal O}(\vec p^2)$. 

In case of vector exchange, only $\Gamma = \gamma^0$ contributes to
${\cal O}(\vec p^{\,0})$ \cite{Iengo2009}. This then again requires $s=s'$
in Eq.(\ref{en5}), giving $A_{\rm res,\ vector} = 4 m_\chi^2$. 
Remembering the additional minus sign in the spin--1 propagator we
therefore again find that vector boson exchange contributes the same
way as scalar exchange. 

Finally, for axial vector exchange, only $\Gamma = \gamma_i \gamma_5 \
(i=1,2,3)$ contribute to ${\cal O}(\vec{p}^{\,0})$. In this case $A_{\rm
  res}$ receives non--vanishing contributions both from $s = s'$ and
from $s = -s'$, so that $A_{\rm res,\ axial \ vector} = - 12
m_\chi^2$. Again including the minus sign from the propagator, we
reproduce our earlier result that axial vector exchange differs from
scalar or vector exchange by a factor of $-3$. Note also that axial
vector exchange describes an interaction between the spins of the two
WIMPs. Such interactions are not suppressed at low velocities.

Now let us discuss $P-$wave annihilation, which corresponds to $n_L =
0$ in Eq.(\ref{en2}). In this case setting $\vec{q} = 0$ would lead to
a result which is of ${\cal O}(\vec{p}^2)$, i.e. of {\em second} order
in the non--relativistic expansion. The leading term is the one linear
in $\vec{q}$; the numerator of Eq.(\ref{en2}) then becomes $-4 m_\chi
\vec{q} \cdot \vec{\gamma}$. Note that the $\gamma_i$ again allow to
access the large component of $\bar{v}(p_2)$. The proper form of the
correction term can then most easily be obtained using trace
techniques, by dividing the 1--loop correction term $\delta A_P
A_{0,P}^\dagger$ by the tree--level result $\left| A_{0,P}
\right|^2$. We find that the $P-$wave correction term differs from the
$S-$wave term by a factor $\vec{p} \cdot \vec{q} / \vec{p}^2$ inside
the momentum integral. Performing the angular integrals, we can write
this in the form of Eq.(\ref{eq:A_S}), with a new function describing
the correction for annihilation from a $P-$wave initial state:
\begin{equation} \label{eq:I_P}
I_P = \Re{\rm e} \left\{ \int_0^\infty dx \frac{2x^2}{x^2-1} \left[ -1 + \frac
  {x^2+1+r}{4x} \ln{ \frac {(x+1)^2+r} {(x-1)^2+r } } \right] \right\}\,,
\end{equation}
where $r = \mu^2 / \vec{p}^2$ as above.

In order to treat the exchange of other bosons, non--trivial Dirac
structures $\Gamma, \ \bar{\Gamma}$ again have to be introduced in
Eq.(\ref{en2}), now for the case $n_L = 0$. Proceeding as above, we
find that pseudoscalar exchange does not lead to a large correction,
while vector and axial vector exchange give the {\em same} correction
as the exchange of a scalar boson. At first glance it may seem
surprising that now axial vector exchange gives the same, positive,
contribution; recall that in case of annihilation from an $S-$wave,
axial vector exchange differed by a factor of $-3$. The difference can
be understood from the observation that axial vector exchange leads to
a spin--spin interaction of the form $4 \vec{s}_1 \cdot \vec{s}_2$,
where $\vec{s}_{1,2}$ are the spins of the two WIMPs. This can be
evaluated using $2 \vec{s}_1 \cdot \vec{s}_2 = S^2 - s_1^2 - s_2^2$,
where $\vec{S} = \vec{s}_1 + \vec{s}_2$ is the total spin. In case of
Majorana WIMPs, annihilation from an $S-$wave requires \cite{dn}
$S=0$, leading to $4 \vec{s}_1 \cdot \vec{s}_2 = -3$. For $P-$wave
annihilation, we need \cite{dn} $S=1$, giving $4 \vec{s}_1 \cdot
\vec{s}_2 = 1$. Note the relative factor of $-3$ between these two
results.

At this point a comment on other WIMPs (than Majorana fermions) is in
order. For a Dirac fermion--antifermion pair, there is no strict
correspondence between the total spin and the orbital angular
momentum. There will also be $S-$wave states with $S=1$ (as the
$J/\psi$ family of quarkonia), as well as $P-$wave states with
$S=0$. In this case the proper factor in front of the axial vector
correction will depend on the spin state, i.e. it will no longer be
completely process independent in this case. However, results for
scalar, pseudoscalar and vector exchange are the same as for Majorana
WIMPs.\footnote{Strictly speaking, Majorana fermions do not have
  diagonal vector couplings. They can, however, have vector couplings
  to other Majorana states. Our result is applicable to this situation
  in the limit where the mass of the second Majorana fermion
  approaches that of the annihilating WIMP.} Finally, there is no such
thing as an axial vector coupling to scalars, but our results for
vector exchange apply to scalar WIMPs as well\footnote{A vector
  coupling exists only for a complex scalar, which can carry a
  charge.}. Scalar exchange now involves a trilinear scalar
interaction $f$, which has dimension of mass. Our results can describe
this situation as well, with $g = f / (2 m_\chi)$.

The integrals in Eqs.(\ref{eq:I_S}) and (\ref{eq:I_P}) should be
understood as principal value integrals, in order to treat the pole at
$x=1$. In case $r \gg 1$ the integrals can be computed analytically,
by expanding the logarithms in inverse powers of $r$. Moreover, in the
limit $r \rightarrow 0$ both $I_S$ and $I_P$ approach $\pi^2/2$,
thereby reproducing the well--known result \cite{LL} that the
one--loop ``Sommerfeld factor'' for massless boson exchange is the
same for $S-$ and $P-$partial waves. We also found accurate numerical
expressions for small and moderate $r$. Altogether, the correction
factors can be described by
\begin{eqnarray} \label{approximation_I}
I_S(r) \simeq \left\{ \begin{array}{ll} \frac{2\pi} {\sqrt{r+1}}
    \left( 1 - \frac {1} {r+2} \right) &  ({\rm large} \ r)\\
\frac{\pi^2/2} { 1 + \frac{ \sqrt{r} } {\pi} + \frac {r}{\pi^2} } &
({\rm small} \ r)  \end{array} \right. \nonumber \\
I_P(r) \simeq \left\{ \begin{array}{ll} \frac {2\pi} {3 \sqrt{r+1}}
    \left( 1 + \frac {1.3} {r+1} \right)&  ({\rm large} \ r) \\
\frac {\pi^2/2} { 1 + \frac {3\sqrt{r}} {\pi} + \frac{r} {\pi} } &
({\rm small} \ r)  \end{array} \right.
\end{eqnarray}
The approximations for large and small $r$ intersect at $r \simeq 5.6
\ (4.2)$ for $I_S \ (I_P)$; the two approximations for $I_P$ intersect
a second time at $r \simeq 6.3$. 

\begin{figure}[h!]
\centerline{\rotatebox{270}{\includegraphics[width=110mm]{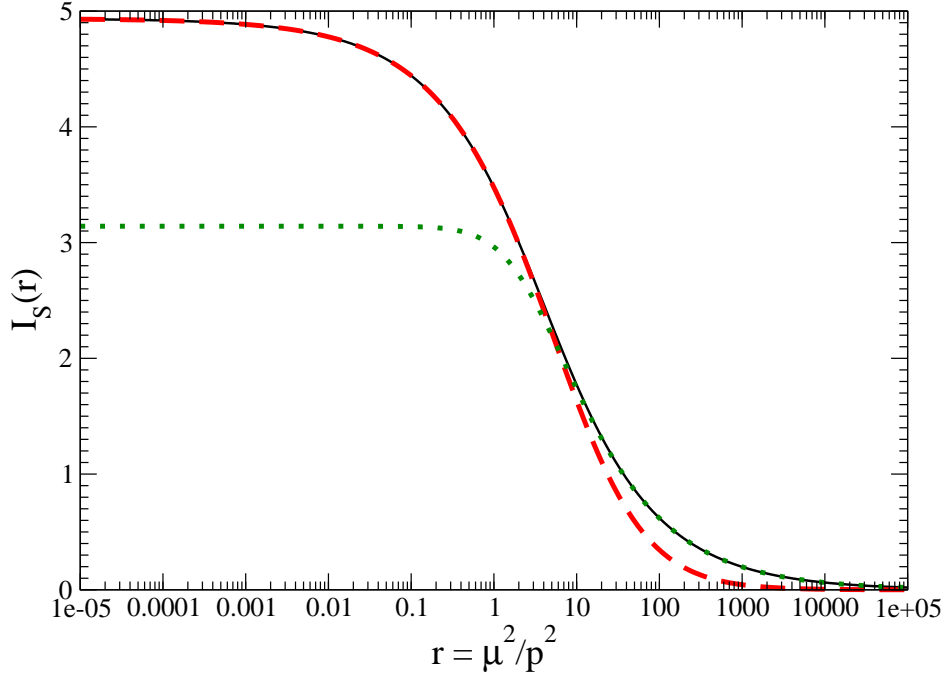}}}
\caption{Comparison between the exact $I_S$ (solid, black) and its
  approximations for large $r$ (dotted, green) and small $r$ (dashed, red).}
\label{fig:approximation.eps}
\end{figure}

In Fig.~\ref{fig:approximation.eps}, the exact function $I_S$ and its
approximations are shown. By switching from the low$-r$ to the
high$-r$ expression at the intersection point one reproduces the exact
numerical result to better than 4\% for all values of $r$. In case of
$I_P$ (not shown) the large$-r$ approximation overshoots rather than
undershoots the exact result for $10^{-3} \lsim r \lsim 1$. This leads
to slightly larger discrepancies, of up to 6\%, between the exact
$I_P$ and its approximations at $r\sim 5$ where the two approximations
intersect. For the purpose of calculating the WIMP relic density a
relative error on the {\em correction} of 6\% is quite acceptable.

\begin{figure}[h!]
\centerline{\rotatebox{270}{\includegraphics[width=110mm]{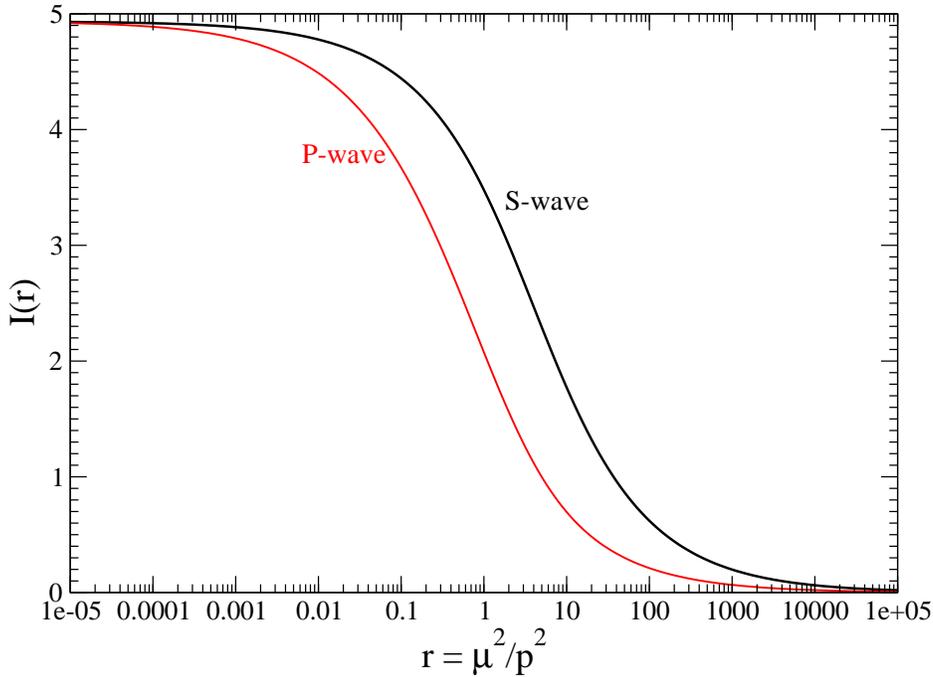}}}
\caption{Comparison between $I_S$ describing the correction for
  annihilation from an $S-$wave (black) and $I_P$ describing the
  correction for annihilation from a $P-$wave (red or grey).}
\label{fig:S_P.eps}
\end{figure}

In Fig.~\ref{fig:S_P.eps} we show $I_S(r)$ and $I_P(r)$ as black and
red (grey) lines, respectively. As noted earlier, the two functions
coincide for massless exchange bosons, $r = 0$. For nonvanishing boson
mass, $r > 0$, the $S-$wave contribution is larger than the $P-$wave
one, by up to a factor of 3 at large $r$; see
Eqs.(\ref{approximation_I}). Note that for any finite mass of the
exchanged boson, $\mu \neq 0$, the zero--velocity limit $|\vec p|
\rightarrow 0$ corresponds to $r \rightarrow
\infty$. Eqs.(\ref{approximation_I}) show that asymptotically
$I_{S,P}(r \rightarrow \infty) \propto 1/\sqrt{r} =
|\vec{p}|/\mu$. Eq.(\ref{eq:A_S}) then shows that the corrections
approach constant values of order $g^2 m_\chi/(4\pi\mu)$ for
$|\vec{p}| \rightarrow 0$, if $\mu \neq 0$. Such corrections will
threaten the convergence of perturbation theory only if the WIMP and
boson masses differ by a loop factor; in the technically more natural
case where the boson mass lies a factor of a few below that of the
WIMP, we still find a significant enhancement relative to the naive
expectation that corrections should be of order $g^2/(16\pi^2)$. We
finally note that our numerical results are consistent with those in
Ref.~\cite{Cassel2009}.

\section{Dark Matter relic density}
\label{sec:relicdensity}

In this section we describe how to calculate the loop--corrected WIMP
relic density from the loop--corrected WIMP annihilation cross section.

The evolution of the WIMP number density $n_\chi$ with time $t$ in the
early Universe is governed by the Boltzmann equation \cite{kt}
\begin{equation} \label{boltz}
\frac {d n_\chi} {dt} + 3 H n_\chi = -\langle \sigma v \rangle \left(
  n_\chi^2 - n_{\chi,{\rm eq}}^2 \right)\,.
\end{equation}
Here $H$ is the Hubble parameter describing the expansion of the
Universe, $\sigma$ is the total WIMP annihilation cross section, $v$
is again the relative velocity between the annihilating WIMPs in their
center of mass frame, $n_{\chi,{\rm eq}}$ is the WIMP number density
in thermal equilibrium, and $\langle \dots \rangle$ denotes thermal
averaging. For non--relativistic kinematics, the latter is given by
\begin{equation} \label{eq:sigmav_av}
\langle \sigma v\rangle = \frac{2x^{3/2}}{\sqrt{\pi}}\int_0^{\infty}
(\sigma v) \frac{v^2} {4} e^{-xv^2/4} dv \,,
\end{equation}
where $x=m_\chi/T$, $T$ being the temperature of the thermal bath.

As long as the WIMP annihilation (or creation) rate is larger than the
Hubble expansion rate, the WIMPs are (almost) in thermal
equilibrium. However, once the annihilation rate falls below the
expansion rate, WIMPs nearly decouple. Their present relic density is
then to very good approximation given by \cite{kt}
\begin{equation} \label{Omega}
\Omega_\chi h^2 = \frac{8.5 \times 10^{-11}~x_F~{\rm GeV}^{-2}} {\sqrt{g_*(x_F)}
    J(x_F)}\,.
\end{equation}
Here $x_F = m_\chi / T_F$, where $T_F$ is the freeze--out temperature
of the WIMPs, $g_*$ is the number of relativistic degrees of freedom,
and the annihilation integral $J(x_F)$ is defined as \cite{kt}
\begin{equation} \label{J}
J(x_F) = \int_{x_F}^\infty \!\! dx \frac{\langle \sigma v
  \rangle}{x^2}\, . 
\end{equation}

The freeze--out temperature of typical WIMPs is rather small, $T_F
\simeq m_\chi/20$, hence WIMPs are non--relativistic when they freeze out.
This suggests an expansion of $\sigma_0 v$ in powers $v$ \cite{kt}:
\begin{equation} \label{eq:sigmav_exp}
\sigma_0 v\simeq \mathscr{A} + \mathscr{B} v^2 + \cdots\,,
\end{equation}
where $\mathscr{A}$ and $\mathscr{B}$ are independent of $v$.
Note that $\mathscr{A}$ contains only $S-$wave contributions, while
$\mathscr{B}$ contains both $S-$ and $P-$wave contributions. While
this often \cite{Griest1991} works rather well for the tree--level
cross section $\sigma_0$, the loop corrections we computed in the
previous Section cannot be parameterized in this way. We saw in
Fig.~\ref{fig:S_P.eps} that the correction factors $I_S$ and $I_P$ depend
strongly on $v$ via the quantity $\sqrt{r} = \mu / |\vec{p}| = 2 \mu /
(m_\chi v)$. We thus have to re--compute $\langle \sigma v \rangle$,
which can then be used in Eq.(\ref{J}) to derive the Dark Matter relic
density using Eq.(\ref{Omega}).

The 1--loop corrected WIMP annihilation cross section for partial wave
labeled by $L$ can be written as
\begin{equation}
\sigma_L = \sigma_{0,L} + \delta\sigma_L\,.
\end{equation}
Eq.(\ref{eq:A_S}) and the analogous expression for annihilation from
the $P-$wave imply that
\begin{eqnarray} \label{del_sig}
\delta\sigma_S &=& \frac{g^2}{2\pi^2v} I_S(r) \sigma_{0,S} \,; \nonumber\\
\delta\sigma_P &=& \frac{g^2}{2\pi^2v} I_P(r) \sigma_{0,P} \, .
\end{eqnarray}
The thermal average over the tree--level cross section, expanded
according to Eq.(\ref{eq:sigmav_exp}), can be computed easily
\cite{kt}:
\begin{equation} \label{sig_0_av}
\langle \sigma_0 v \rangle(x) \simeq \mathscr{A} + \frac{6
  \mathscr{B}} {x} \,.
\end{equation}
In order to calculate the thermal averages over the correction terms,
we rewrite the integral in Eq.(\ref{eq:sigmav_av}) in terms of the
integration variable $t = v \sqrt{x}$:
\begin{equation} \label{en10}
\langle \delta \sigma_L v \rangle = \frac {1}{2 \sqrt{\pi}} \int_0^\infty
dt \, t^2 (\delta \sigma_L v) {\rm e}^{-t^2/4}\,.
\end{equation}
Eqs.(\ref{del_sig}) imply that the $t-$dependence of $\delta \sigma_L
v$ takes the form $v^n I_L(r) = (t / \sqrt{x})^n I_L \left(\frac{4 \mu^2 x}
{m_\chi^2 t^2} \right)$, with $n = -1 \ (+1)$ for annihilation from
an $S- \ (P-)$wave initial state. This implies that the thermal
average over the correction terms to the WIMP annihilation cross
section can be written as $x^{-n/2}$ times a function of $z = 2 \mu
\sqrt{x} / m_\chi$.

In Appendix A we describe how we parameterize the resulting
functions for $S-$ and $P-$wave annihilation. The one--loop
corrections to the ``annihilation integrals'' can then be written as
\begin{eqnarray} \label{eq:Jxf}
\delta J_S(x_f) &=& \frac{g^2\mathscr{A}} {\pi^{5/2}} \frac {\mu} {m_\chi}
\int_{z_F}^{\infty} \frac {dz} {z^2} \left( \frac{1} {a_Sz^2 + b_Sz + c_S} +
  d_S \right) \,, \nonumber \\
\newline
\delta J_P(x_f) &=& \frac {64 g^2 \mathscr{B} } {\pi^{5/2}} \left(
  \frac {\mu} {m_\chi} \right)^3 \int_{z_F}^{\infty} \frac {dz} {z^4}
\left[ \exp(-a_Pz + b_P) + \frac{1}{c_P z + d_P} \right]\, . 
\end{eqnarray}
Here $z_F = (2 \mu / m_\chi) \sqrt{x_F}$, and the coefficients $a_S,
\, b_S, \, c_S, \, d_S$ and $a_P,\, b_P,\, c_P,\, d_P$ are given in
Eqs.(\ref{eq:ABCF}) and (\ref{eq:ABCDEF}), respectively. We have made
the simplifying assumption that $\mathscr{B}$ is dominated by $P-$wave
contributions. This is true whenever the contribution $\propto
\mathscr{B}$ to the annihilation integral is comparable to, or
dominates over, the one $\propto \mathscr{A}$. In the opposite case
the {\em correction} to the $\mathscr{B}-$term will in any case be
insignificant.\footnote{The one--loop correction to the ${\cal
    O}(v^2)$ contribution to the $S-$wave annihilation cross section
  involves the sum of two terms: the product of the ${\cal O}(v^0)$
  one--loop correction and the ${\cal O}(v^2)$ tree--level amplitude,
  and the product of the ${\cal O}(v^2)$ one--loop correction and the
  ${\cal O}(v^0)$ tree--level amplitude. Only the first of these terms
  can be computed using the results of Sec.~2.}

Not surprisingly, the corrections are quadratic in the coupling of the
exchanged boson to the WIMP. Moreover, they depend on the boson mass
{\em only} through the ratio $\mu/m_\chi$. Recall finally that $I_S$
has to be multiplied with $-3$ if the exchanged boson has axial vector
couplings; the same factor has to be included in $\delta J_S(x_F)$.

\begin{figure}[ht]
\centerline{\includegraphics[keepaspectratio=true,width=130mm]{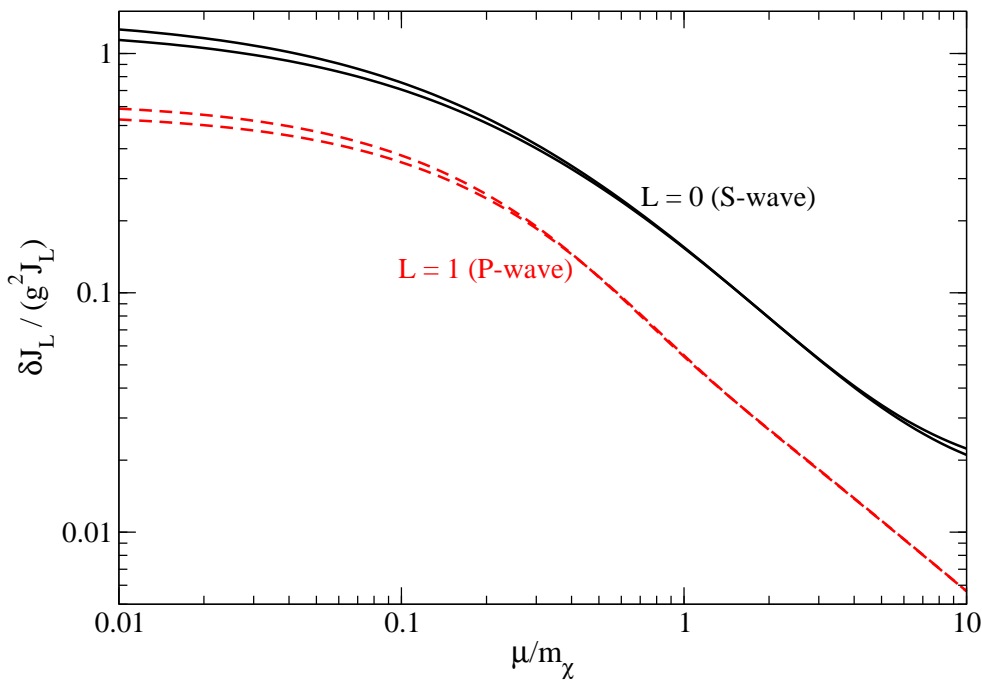}}
\vspace*{1mm}
\caption{Corrections to the annihilation integrals for annihilation
  from the $S-$ wave (solid, black) and $P-$wave (dashed, red or
  grey), normalized to the tree--level results, and divided by the
  square of the WIMP coupling to the exchanged boson $g$. The upper
  (lower) curves are for inverse scaled freeze--out temperature $x_F =
  m_\chi / T_F = 25 \ (20)$.}
\label{fig:delj}
\end{figure}

In Fig.~\ref{fig:delj} we show the {\em relative} size of the loop
corrections to the annihilation integrals for $S-$ and $P-$wave
annihilation, divided by the square of the WIMP--boson coupling
$g$. Eq.(\ref{Omega}) shows that this also gives the relative change
of the relic density due to our loop corrections, as long as the
corrections are small, in which case $x_F$ remains essentially
unchanged by these corrections; note that $x_F$ depends only
logarithmically on the annihilation cross section. This is the main
model--independent result of our paper. The size of the corrections to
the relic density can be read off directly, by simply inserting the
values of masses $\mu$ and $m_\chi$ and coupling $g$ given in a
concrete model.

We see that the corrections are less important for $P-$wave
annihilation. This is true even for $\mu \ll m_\chi$, where the loop
functions $I_S$ and $I_P$ become equal. In this limit $v \delta
\sigma_S \propto 1/v$ while $v \delta \sigma_P \propto v$. Performing
the integrals for the thermal averaging, Eq.(\ref{eq:sigmav_av}), and
inserting the results into the definition of the annihilation
integral, Eq.(\ref{J}), one finds
\begin{equation} \label{deljrat}
\frac {\delta J_P / J_P} { \delta J_S / J_S} = \frac {4}{9} \ \ \ {\rm
  for} \ \mu \rightarrow 0\,.
\end{equation}
The ratio becomes even smaller for nonvanishing $\mu$, because then
$I_P < I_S$.

Fig.~\ref{fig:delj} shows results for $x_F = 25$ (upper curves) and
$x_F = 20$ (lower curves); this spans the range of decoupling
temperatures in usual WIMP models. One can easily show analytically
that in the limit $\mu \rightarrow 0$, $\delta J_L / J_L \propto
\sqrt{x_F}$ for both $L=0$ ($S-$wave) \cite{baro} and $L=1$
($P-$wave). On the other hand, Fig.~\ref{fig:delj} shows that the {\em
  relative} correction to the annihilation integral becomes
independent of $x_F$ once $\mu \gsim 0.3 m_\chi$. This figure also
shows that for the most plausible scenarios with electroweak strength
couplings, possibly suppressed by mixing effects, where $g^2 \lsim
0.5$, the loop corrections are significant only for $\mu \lsim
m_\chi$, as stated in the beginning of Sec.~2.

Two comments are in order before concluding this Section. First,
Fig.~\ref{fig:delj} seems to imply that, especially for $P-$wave
annihilation, the corrections are never very large. This is
misleading. A simple one--loop calculation can be trusted only if
$\delta \sigma < \sigma$ for {\em all} relevant velocities $v$. This
requires $g^2 I_{S,P} / (v \pi^2) < 1$. For $\mu \rightarrow 0$ this
will always be violated at sufficiently small $v$, requiring summation
of higher orders. In the standard treatment
\cite{summing,Iengo2009,Cassel2009} this leads to $\delta \sigma
\propto 1/v$ even for annihilation from a $P-$wave initial state. We
saw at the end of Sec.~2 that for finite $\mu$ the maximal size of the
correction to the cross section is of relative size $g^2 m_\chi / (4 \pi
\mu)$. Our one--loop calculation can be trusted as long as this
quantity is well below 1.

Secondly, it has very recently been pointed out \cite{white} that
Eq.(\ref{J}) becomes inadequate for very small $\mu$. In this case the
annihilation integral receives sizable contributions from quite large
$x$, i.e. from low temperatures. Eq.(\ref{J}) assumes that the WIMPs
are in kinetic equilibrium while they are annihilating. It has to be
modified for temperatures below the kinetic decoupling temperature,
which is typically a few (tens of) MeV \cite{kinetic}. This
modification can have sizable effects for very small $\mu$
\cite{white}.  However, we just saw that our strictly perturbative
treatment is not reliable in this case anyway. Recall that $v \delta
\sigma$ becomes constant, rather than scaling like $1/v$, for
velocities below $v_{\rm crit} = \mu/m_\chi$. Our perturbative
treatment will be reliable only if $\mu / m_\chi \gg g^2/(4\pi) \sim
0.01$. This implies that WIMP annihilation will quickly become
irrelevant for $x > 1/v^2_{\rm crit} \sim 10^4$, well before kinetic
decoupling occurs.


\section{Applications}
\label{sec:numerical_calculation}

In this Section we apply our results to existing WIMP models. We start
with two simple models with scalar or fermionic SM singlets; in the
third Subsection we discuss the more widely studied case of the MSSM
neutralino. 

\subsection{Scalar singlet WIMP}

This is the simplest WIMP model \cite{simplest}. One only needs to
introduce a single real scalar field $\chi$ to describe Dark
Matter. If one forbids terms linear in $\chi$ by some (possibly
discrete) symmetry in order to prevent $\chi$ decays, the only
renormalizable coupling to SM fields allowed by all symmetries is of
the form $\chi^2 |h|^2$, where $h$ is the scalar Higgs doublet. Upon
weak symmetry breaking this gives rise to a trilinear scalar
interaction of the form $V \chi^2 \phi$, where $\phi$ is the physical
Higgs scalar of the SM and $V = 246$ GeV the vacuum expectation value
(vev) of the Higgs.

These interactions allow $\chi$ to annihilate via $\phi$ exchange in
the $s-$channel; annihilation into two $\phi$ bosons is also allowed
for $m_\chi > m_\phi$. An accurate tree--level calculation of the
resulting relic density has been performed in \cite{scalar_rel}.
Writing the coefficient of the $\chi^2 |h|^2$ term in the Lagrangian
as $-k/2$, they find that the correct relic density (\ref{range}) is
obtained for $k \simeq 0.28 m_\chi / (1 \ {\rm TeV})$, unless $m_\chi
\sim m_\phi / 2$, in which case an even smaller $k$ is required. We
can use our formalism to compute corrections to this result from
$\phi$ exchange prior to annihilation, i.e. $\varphi = \phi$ in this
case. This gives a coupling factor\footnote{Recall from our discussion
  near the end of Sec.~2 that the relevant quantity for a purely
  scalar theory is the trilinear scalar coupling divided by $2
  m_\chi$, i.e. in the case at hand, $g = k V / (2 m_\chi)$.} $k^2 V^2
/ (4 m_\chi^2) \simeq 0.0012$, (almost) independent of
$m_\chi$. Fig.~\ref{fig:delj} shows that the corrections due to $\phi$
exchange in the initial state will then at best be at the permille
level. In this model we therefore do not find any significant
radiative corrections involving the initial state only. This should
also hold for the inert doublet model \cite{inert}, since the coupling
between the inert WIMP doublet and the SM Higgs boson will have to
satisfy a similar relation.

\subsection{Fermionic singlet WIMP}

The next simplest WIMP model \cite{singlet} contains a Dirac fermion
SM singlet $\chi$ as well as a real scalar singlet $\varphi$, with
couplings $g \bar \chi \chi \varphi + A \varphi |h|^2$, where $h$ is
again the Higgs doublet of the SM. The latter term induces mixing
between the singlet $\varphi$ and the SM Higgs boson, allowing
$\varphi$ to decay. If this mixing is small and $m_\varphi < m_\chi$,
the dominant $\bar\chi \chi$ annihilation channel is into two
$\varphi$ bosons, via $t-$ and $u-$channel diagrams. In the
non--relativistic limit this is a pure $P-$wave process, with
tree--level cross section
\begin{equation}\label{sig_sing}
v \sigma(\bar\chi\chi \rightarrow \varphi \varphi) = \frac {g^4 v^2
  \beta} {24 \pi} \frac {m_\chi^2 \left( 9 m_\chi^4 - 8 m_\chi^2
    m_\varphi^2 + 2 m_\varphi^4 \right) } { \left( 2 m_\chi^2 -
    m_\varphi^2 \right)^4} + {\cal O}(v^4)\,,
\end{equation}
where $\beta = \sqrt{1-m_\varphi^2/m_\chi^2}$. In the limit $m_\chi^2
\gg m_\varphi^2$ this simplifies to 
$$v \sigma(\bar\chi\chi \rightarrow \varphi \varphi) = \frac {3 g^4
  v^2} {128 \pi m_\chi^2}\, .$$
The tree--level calculation therefore predicts the correct relic
density (\ref{range}) for coupling 
$$g^2 \simeq 0.2 \frac {m_\chi} {100 \ {\rm GeV}} \,.$$

\vspace*{-6mm}
\begin{figure}[ht]
\centerline{\rotatebox{270}{\includegraphics[keepaspectratio=true,width=130mm]{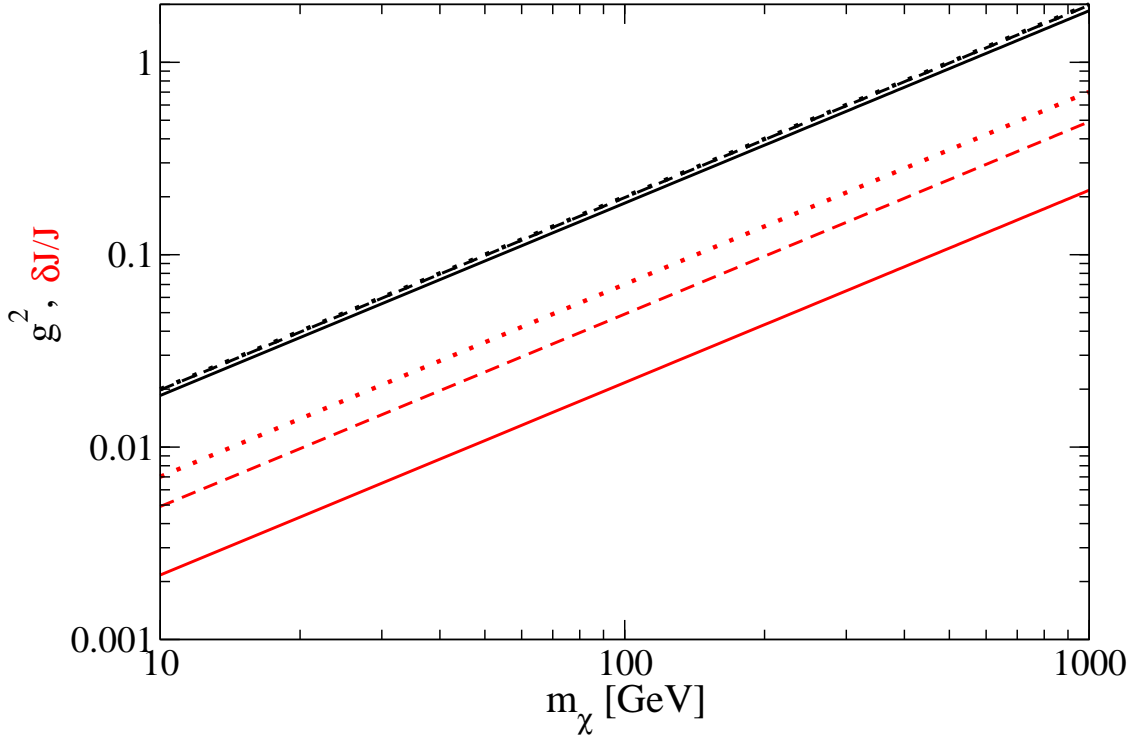}}}
\vspace*{-3mm}
\caption{Strength of the $\bar \chi \chi \varphi$ coupling $g$ (black),
as well as the resulting one--loop correction to the annihilation
integral (red), in a model with Dirac fermion singlet DM. $g$ has been
computed from the requirement that $\bar \chi \chi \rightarrow \varphi
\varphi$ annihilation produces the correct $\chi$ relic density at
tree level. The solid (dashed, dotted) curves are for $\mu/m_\chi =
0.5 \ (0.2,\ 0.1)$, where $\mu \equiv m_\varphi$.}
\label{fig:singlet}
\end{figure}

This can then be used to read off the correction due to $\varphi$
exchange in the initial state from Fig.~\ref{fig:delj}, with $\mu =
m_\varphi$. The result is shown in Fig.~\ref{fig:singlet}, where we
have used the exact (tree--level) result (\ref{sig_sing}) to derive
the required coupling strength. The black curves show that this
coupling strength depends only very weakly on the mass of the scalar
for $\mu \equiv m_\varphi \lsim 0.5 m_\chi$. Note that the cross
section slightly {\em increases} with increasing $m_\varphi$ as long
as $m_\varphi < 0.85 m_\chi$. The reason is that increasing
$m_\varphi$ allows the $t-$ and $u-$channel propagators to be less
off--shell, as shown by the denominator in Eq.(\ref{sig_sing}). As a
result, the required coupling strength slightly decreases with
increasing $m_\varphi$.

However, the red curves show that this effect is much smaller than the
strong dependence of $\delta J_P/J_P$ on $\mu/m_\chi$ illustrated in
Fig.~\ref{fig:delj}. More importantly, we see that the corrections due
to $\varphi$ exchange in the initial state can easily exceed the
uncertainty of the observational determination of the DM relic
density; for $\mu/m_\chi = 0.5 \ (0.2,\ 0.1)$ they even reach the 10\%
level for $m_\chi > 450 \ (210, \ 130)$ GeV. Since the correction is
positive, one would have to reduce the coupling in order to obtain the
correct relic density after inclusion of one--loop corrections. This
would correspondingly reduce all interactions between the WIMP $\chi$
and the SM particles, all of which are mediated by $\varphi$ exchange.

\subsection{The lightest neutralino in the MSSM}

We finally want to apply our formalism to the lightest neutralino in the
minimal supersymmetric extension of the Standard Model (MSSM), which
is the probably best motivated WIMP, and certainly the most widely
studied \cite{Silk_rev,jungman} one. For simplicity we will assume
that sfermions are heavy. Given experimental lower bounds on the
masses of sfermions and Higgs bosons, relatively light sfermions by
themselves typically only lead to an acceptable neutralino relic
density in the presence of significant co--annihilation
\cite{sfer_coan}. This involves several particles, with mass
splittings of order of the absolute value of the 3--momentum in the
initial state. These more complicated scenarios cannot be treated with
the formalism presented in this paper.\footnote{Very recently the
  summation of ``Sommerfeld corrections'' for the case of nearly
  degenerate states was discussed in \cite{generalized}.}

Generally speaking, in the MSSM the neutralinos are mixtures of the
$U(1)_Y$ gaugino $\tilde B$, the neutral $SU(2)$ gaugino $\widetilde{W}_3$,
and of the two neutral higgsinos $\tilde h_1^0, \ \tilde h_2^0$:
\begin{equation} \label{en20}
\tilde \chi^0_i = N_{i1} \tilde{B} + N_{i2} \widetilde{W}_3 + N_{i3}
\tilde{h}_1^0 + N_{i4} \tilde{h}_2^0 \,\ (i=1, \cdots, 4).
\end{equation}
The coefficients $N_{ik}$ satisfy the sum rule $\sum_{k=1}^4 \left|
  N_{ik} \right|^2 = 1 \ \forall i$.  Most phenomenological analyses
of the MSSM assume that the soft SUSY breaking gaugino masses unify at
or near the scale of Grand Unification \cite{jungman}. This implies
that the $U(1)_Y$ gaugino mass is about half the $SU(2)$ gaugino mass
near the TeV scale. As a result, the Wino component of our candidate
WIMP, the lightest neutralino ($\chi \equiv \tilde \chi_1^0$), is
subdominant, i.e. $\left| N_{11} \right|^2 \gg \left| N_{12}
\right|^2$. If sfermions are heavy, $\chi$ annihilation involves
couplings of the lightest neutralino to gauge or Higgs bosons, which
vanish in the pure Bino limit ($\left| N_{11} \right| \rightarrow
1$). In models with gaugino mass unification and heavy sfermions, the
annihilation cross section can thus only be sufficiently large if
$\chi$ has significant higgsino components.

On the other hand, for a nearly pure higgsino, where $\left| N_{11}
\right|^2 + \left| N_{12} \right|^2 \ll 1$, the cross section for
annihilation into $W^+W^-$ and $Z^0 Z^0$ pairs is so large that a mass
$m_\chi \simeq 1$ TeV is required to obtain the correct relic density
(\ref{range}) \cite{swedes}. Such a large mass for the lightest
superparticle is at odds with the primary motivation for postulating
the existence of superparticles, which is the stabilization of the
weak scale against quadratically divergent quantum corrections.

Assuming gaugino mass unification, the most natural neutralino
satisfying the constraint (\ref{range}) is therefore a bino--higgsino
mixture, dubbed a ``well--tempered neutralino'' in
ref.\cite{tempered}. Note that the neutralino couplings to Higgs
bosons involves products of a combination of gaugino components
($N_{i2} - \tan \theta_W N_{i1}$, where $\theta_W$ is the weak mixing
angle) with one of the higgsino components ($N_{i3}, \ N_{i4}$). These
couplings are maximal in the region of strong gaugino--higgsino
mixing, and should thus be sizable for the ``well--tempered''
neutralino. Moreover, as well known, at least one of the neutral MSSM
Higgs bosons is rather light, with mass below 130 GeV. This leads one
to expect potentially sizable 1--loop corrections due to the exchange
of a Higgs boson prior to WIMP annihilation.

We checked this with the help of the the code \texttt{micrOMEGAs 2.2}
\cite{micromegas}. Among other things, this program computes the
complete tree--level neutralino annihilation cross sections for all
two--body final states. It does not resort to the non--relativistic
expansion (\ref{eq:sigmav_exp}), but one can easily determine the
coefficients $\mathscr{A}$ and $\mathscr{B}$ by calculating the
annihilation cross section at two different values of $v$, i.e. for two
(slightly) different cms energies $\sqrt{s}$. These coefficients are
then used in Eqs.(\ref{eq:Jxf}) to compute the {\em corrections} to
the annihilation integral; we emphasize that we continue to use the
full cross sections, not their non--relativistic expansions, for the
calculation of the tree--level contribution to the annihilation
integral. We also take $x_F$ from the program. The one--loop
corrected $\chi$ relic density can then be expressed as
\begin{equation} \label{en21}
\Omega_\chi h^2 = \Omega_{\chi,0} h^2 \frac {J_0(x_F)} {J_0(x_F) +
  \delta J(x_F)}\, 
\end{equation}
where $\delta J = \delta J_S + \delta J_P$, and the tree--level value
$J_0(x_F)$ can be calculated from the program's tree--level prediction
$\Omega_{\chi,0} h^2$ using Eq.(\ref{Omega}).

\begin{figure}[h!]
\centerline{\rotatebox{270}{\includegraphics[width=110mm]{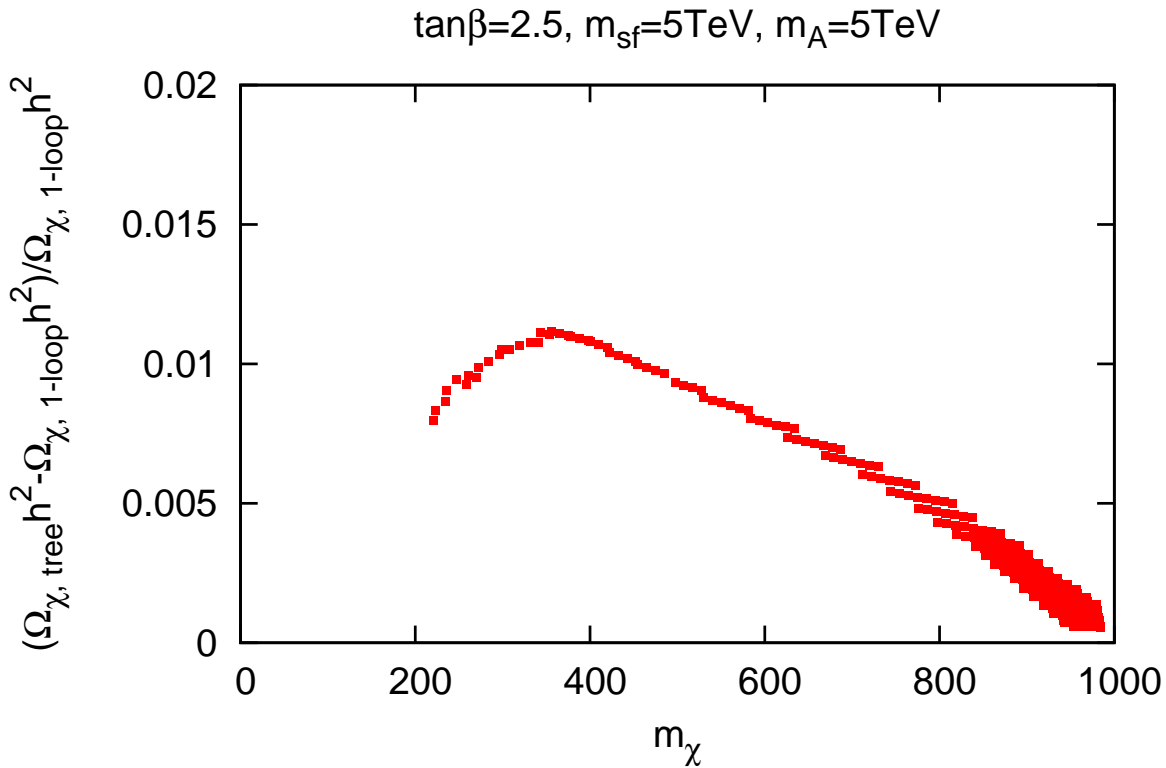}}}
\vspace*{2mm}
\caption{The relative size of one-loop correction to the relic density
  of a ``well-tempered'' neutralino. The ratio between $M_1$ and $\mu$
  is fixed such that the relic density satisfies the constraint
  (\ref{range}) within two standard deviations. We take $M_2=2M_1$,
  consistent with gaugino mass unification, $\tan\beta = 2.5$, and
  assume sfermions and most Higgs bosons to be very heavy. Scenarios
  with $m_\chi <200$ GeV are excluded by Higgs boson searches.}
\label{fig:welltempered}
\end{figure}

The result is shown in Fig.~\ref{fig:welltempered}. It has been
generated using micrOMEGAs \cite{micromegas}, using Softsusy
\cite{softsusy} to calculate the superparticle and Higgs boson
spectrum; we specify the input directly at the weak scale. All points
satisfy the relic density constraint (\ref{range}) within two standard
deviations. For simplicity we take (unnaturally) large values for the
masses of all sfermions and of the CP--odd Higgs boson $A$, but the
results would not change significantly as long as $m_{\rm sf} \gsim 2
m_\chi$ and $m_A \gsim 3 m_\chi$. We include loop corrections due to
exchange of the $Z-$boson as well as both CP--even neutral Higgs
boson, but the contribution from the heavier Higgs boson is totally
negligible due to its large mass.

We see that the corrections are most important for $m_\chi$ near 350
GeV. For smaller WIMP mass the corrections are reduced because the
higgsino component of $\chi$ becomes smaller, and because the ratio of
light Higgs and WIMP masses becomes smaller, which reduces the loop
functions. The latter effect would tend to increase the correction for
heavier WIMPs. However, for $m_\chi > 350$ GeV the gaugino components
of $\chi$ decrease quickly; this reduces its coupling to the light
Higgs boson, which is most important here. Moreover, for $m_\chi \gsim
900$ GeV, co--annihilation with $\tilde\chi_1^\pm$ and $\tilde
\chi_2^0$ become important \cite{dn,oldcoan,swedes}. The effect of light
boson exchange corrections to co--annihilation is beyond the scope of
our paper, and has not been included in Fig.~\ref{fig:welltempered}.
As a result, the correction becomes comparable to the anticipated
post--PLANCK precision of the observational determination of
$\Omega_\chi h^2$ only for a rather narrow range of $m_\chi \sim 350$
GeV. This is consistent with the results of Fig.~\ref{fig:delj}, given
the fact that the coupling of our ``well--tempered'' neutralino to the
lightest Higgs boson does not exceed 0.2.

The small size of the corrections due to boson exchange in the initial
state indicate that these may well {\em not} be the leading radiative
corrections in the MSSM. In fact, full electroweak one--loop calculations
\cite{baro,boudjema} found much larger corrections in some cases. These are
presumably due to UV--sensitive effects, which cannot be treated using
our formalism. Moreover, QCD corrections can significantly affect the
annihilation cross section into quarks \cite{baro,freitas}.

\section{Summary and Discussion}

We have calculated one--loop corrections to the WIMP annihilation
cross section, and the corresponding corrections to the relic
density, due to the exchange of a relatively light boson between the
WIMPs.

The formalism for calculating the corrections to the cross section is
described in Sec.~2. We used a non--relativistic formalism
\cite{itz,Iengo2009}, i.e. we only included small loop momenta. The
correction can then also be understood as re--scattering of the WIMPs
prior to their annihilation. The motivation for this is that these
configurations can give rise to sizable corrections if the exchanged
boson is lighter than the WIMP and the relevant coupling is not too
small. Note that these corrections are universal, i.e. they are the
same for all final states. However, we saw that they do depend on the
partial wave of the initial state, being smaller for annihilation from
the $P-$wave if the exchanged boson is massive. The main result of
this Section is that these corrections can be described by loop
functions which only depend on the ratio of the mass of the exchanged
boson and the cms three--momentum of the annihilating WIMPs; simple
yet accurate parameterizations for these loop functions are given in
Eqs.(\ref{approximation_I}). We also checked explicitly that the
correction is independent of the form of the coupling of the exchanged
boson, except for the case of axial vector exchange in the $S-$wave,
where an additional factor of $-3$ is required. Moreover, the
corrections can be used for scalar as well as fermionic WIMPs.

In Sec.~3 we computed the resulting correction to the relic
density. We saw that the thermal averaging and the calculation of the
``annihilation integral'' further reduce the relative importance of
the corrections in case of annihilation from a $P-$wave initial
state. The main model--independent result of our paper is shown in
Fig.~\ref{fig:delj}, which shows the relative size of the corrections
to the annihilation integral -- or, almost equivalently, to the relic
density -- as a function of the ratio of the masses of the exchanged
boson and the WIMP. We saw that for ${\cal O}(1)$ coupling between
boson and WIMP, the corrections are significant unless the boson is
heavier than the WIMP. 

Finally, in Sec.~4 we applied this formalism to several WIMP
models. We showed that these corrections are very small for a scalar
singlet WIMP, but can be very large for a Dirac fermion singlet WIMP
annihilating into a light scalar singlet. Finally, we analyzed the
``well--tempered'' neutralino, which is a mixture of $U(1)_Y$ gaugino
and higgsinos, and found corrections comparable to the anticipated
post--PLANCK precision of the observationally determined WIMP relic
density only for a narrow range of neutralino masses near 350 GeV. The
main reason is that the relevant coupling is always below 0.2 in this
case. In a supersymmetric scenario, bigger couplings of WIMPs to Higgs
bosons are e.g. possible for ``singlino'' Dark Matter in scenarios
where the MSSM is extended by an additional Higgs singlet superfield
\cite{singlino}.

In the MSSM diagonal couplings of the lightest neutralino to gauge or
Higgs bosons are always suppressed by mixing angles. Off--diagonal
couplings may be large, however. These can give rise to sizable
corrections if these couplings involve particles that are only
slightly heavier than the lightest neutralino. Such states, with
sizable couplings to the lightest neutralino, often exist in regions
of parameter space where co--annihilation is important. In order to
treat this, one has to extend the formalism presented here to
scenarios where the particles in the loop have slightly different
masses than the annihilating external particles. Moreover, if
co--annihilation with a sfermion is important, there are vertex
corrections involving the exchange of an SM fermion, rather than a
boson, between the co--annihilating particles. This offers another
avenue for future work.

\subsubsection*{Acknowledgment}

We thank A. Pukhov for his help with using micrOMEGAs. KIN is supported
in part by Grants-in-Aid for Scientific Research from the Ministry of
Education, Culture, Sports, Science, and Technology of Japan, and by
the Grant-in-Aid for Nagoya University Global COE Program, ``Quest for
Fundamental Principles in the Universe: from Particles to the Solar
System and the Cosmos'', from the Ministry of Education, Culture,
Sports, Science and Technology of Japan. MD and JMK are partially
supported by the Marie Curie Training Research Networks
``UniverseNet'' under contract no. MRTN-CT-2006-035863, and ``UniLHC''
under contract no. PITN-GA-2009-237920.

\section*{Appendix A: Parameterizations of the thermal averages}

The thermal average of the correction to the $S-$wave annihilation
cross section is given by
\begin{eqnarray} \label{ea1}
\langle \delta \sigma_S v\rangle &=& \frac{x^{3/2}} {2 \pi^{1/2}}
\int_0^\infty v^2 \left( \frac{g^2} {2\pi^2 v} I_S(v) (\sigma_{0,S} v)
\right) {\rm e}^{-xv^2/4} \nonumber \\
&=& \frac{g^2x^{3/2}} {4\pi^{5/2}} \cdot \mathscr{A} \cdot
\int_0^\infty v^2 \left( \frac{I_S(v)}{v} \right) {\rm e}^{-xv^2/4}\,. 
\end{eqnarray}
$\mathscr{A}$ has been defined via the nonrelativistic expansion
of $\sigma v$ in Eq.(\ref{eq:sigmav_exp}). 
The $v-$dependence of $I_S$ can be read off Eq.(\ref{approximation_I}):
\begin{eqnarray} \label{ea2}
I_S(v) \simeq \left\{ \begin{array}{ll} \frac{2\pi v \sqrt{u^2+v^2} }
    {u^2+2v^2} , & v\leq \frac{u}{2.4}\\
\frac{\pi^2/2} {1 + \frac {u} {\pi v} + \frac {u^2} {\pi^2v^2} }, & v
> \frac{u}{2.4}  \end{array} \right. .
\end{eqnarray}
Here we have introduced the quantity $u = 2 \mu / m_\chi$.
We showed in Sec.~3 that the integral in Eq.(\ref{ea1}) is a function
of the variable $z= u \sqrt{x}$. We find the following fitting function
for the ``thermally averaged'' $(I_S/v)$, defined as the integral in
the last line of Eq.(\ref{ea1}):
\begin{eqnarray}
\langle \frac{I_S}{v} \rangle_{\rm fit} = \frac{1}{x} \left(
  \frac{1}{a_S z^2 + b_S z + c_S} + d_S \right), 
\end{eqnarray}
with 
\begin{equation} \label{eq:ABCF}
a_S = 0.000593;\  b_S = 0.03417;\ c_S = 0.1015;\ d_S = 0.1182.
\end{equation}

For the $P-$wave, 
\begin{eqnarray}
\langle \delta \sigma_P v\rangle &=& \frac{x^{3/2}} {2 \pi^{1/2}}
\int_0^\infty v^2 \left( \frac{g^2}{2\pi^2 v} I_P(v)(\sigma_{0,P} v)
\right) {\rm e}^{-xv^2/4} \nonumber\\
&=& \frac {g^2x^{3/2}} {4\pi^{5/2}} \cdot \mathscr{B} \cdot
\int_0^\infty v^2 \left( v^2 \cdot \frac{I_P(v)}{v} \right) {\rm e}^{-xv^2/4}, 
\end{eqnarray}
with 
\begin{eqnarray}
I_P(v) \simeq \left\{ \begin{array}{ll} \frac{2\pi v \left( u^2 +
        2.3v^2 \right) } {3(u^2+v^2)^{3/2}}, &  v\leq \frac{u}{2.1}\\
\frac{\pi^2/2} {1 + \frac{3u}{\pi v} + \frac{u^2}{\pi v^2} }, &
v>\frac{u}{2.1}  \end{array} \right. .
\end{eqnarray}
As above, we find a fitting function for the ``thermally averaged''
$(v I_P)$:
\begin{eqnarray}
\langle v I_P \rangle_{\rm fit} = \frac{16}{x^2} \left( {\rm e}^{-a_P z -
    b_P} + \frac {1} { c_P z + d_P} \right),
\end{eqnarray}
with
\begin{equation}\label{eq:ABCDEF}
a_P = 0.318;\ b_P = 0.1226; c_P = 0.3309; d_P = 0.6306.
\end{equation}

\section*{Appendix B: Comparison to a full one--loop calculation}

In this Appendix we compare our approximate treatment of corrections
due to $\varphi$ exchange with a full one--loop calculation. We do
this in the framework of a purely scalar theory, where the exact
vertex correction is UV finite. As we remarked in Sec.~2, our
formalism will not capture corrections associated with the
renormalization of the coupling(s) relevant for WIMP annihilation, so
chosing an example with UV--finite vertex correction greatly
simplifies the comparison to the full one--loop calculation.

\begin{center}
\begin{picture}(400,180)(0,0)
\DashLine(0,160)(70,90){2}
\DashLine(0,20)(70,90){2}
\Vertex(70,90){4}
\SetColor{Red}
\DashLine(35,125)(35,55){2}
\SetColor{Black}
\Text(27,90)[]{\Red{$\varphi$}}
\Text(0,150)[]{$\chi$}
\Text(0,30)[]{$\chi$}
\Text(55,115)[]{$\chi$}
\Text(55,67)[]{$\chi$}
\Text(45,10)[]{\large a)}
\DashLine(120,160)(190,90){2}
\Vertex(190,90){4}
\DashLine(120,20)(140,40){2}
\DashLine(170,70)(190,90){2}
\DashCArc(155,55)(20,45,225){2}
\SetColor{Red}
\DashCArc(155,55)(20,225,45){2}
\SetColor{Black}
\Text(180,50)[]{\Red{$\varphi$}}
\Text(130,58)[]{$\chi$}
\Text(120,150)[]{$\chi$}
\Text(120,30)[]{$\chi$}
\Text(165,10)[]{\large b)}
\DashLine(240,160)(310,90){2}
\DashLine(240,20)(310,90){2}
\Vertex(310,90){4}
\SetColor{Red}
\DashLine(270,50)(300,50){2}
\SetColor{Black}
\Text(310,50)[]{\Red{$\varphi$}}
\Text(240,150)[]{$\chi$}
\Text(240,30)[]{$\chi$}
\Text(282,73)[]{$\chi$}
\Text(285,10)[]{\large c)}

\end{picture}
\end{center}
\small{Figure 8: Feynman diagrams describing full initial state
  radiative corrections: vertex correction (a), wave function
  renormalization (b) and real emission (c); wave function
  renormalization of, and emission off, the upper leg also has to be
  included. The WIMP $\chi$ and the exchanged boson $\varphi$ are
  denoted by black and red (grey) dashed lines, respectively, while
  the blob denotes the $\chi \chi$ annihilation vertex, which is
  independent of the $\chi$ momenta.}

\setcounter{figure}{8}
\vspace*{5mm}

The Feynman diagrams describing exact one--loop corrections associated
with the initial state are shown in Fig.~8. Here the blob describes
the (tree--level) $\chi$ annihilation process; this could e.g. be a
quartic vertex involving two lighter scalars, or a trilinear vertex
coupling to the $s-$channel propagator of another scalar particle. For
the purpose of our calculation we only need to know that the rest of
the diagram described by the blob is independent of the loop momentum.
We describe the $\varphi \chi \chi$ vertex by the (dimensionful)
coupling $\kappa$.

Let us begin by computing the vertex correction; recall that this is
the only diagram that contributes in the approximate treatment of
Sec.~2. It gives:
\begin{equation} \label{ap1}
\frac{A_{\rm vertex}} {A_0} = i \kappa^2 \int \frac {d^4
  q}{(2\pi)^4} \frac {1}{(P+q)^2 - m^2_\chi} \frac{1}{(P-q)^2 -
  m_\chi^2} \frac{1}{ (p-q)^2 - \mu^2} \,.
\end{equation}
Here $A_0$ is the tree--level matrix element described by the
blob in Fig.~8. Recall that $P = (p_1+p_2)/2, \ p = (p_1-p_2)/2$,
where $p_{1,2}$ are the 4--momenta of the incoming WIMPs, $\mu$ is the
mass of $\varphi$, and $\kappa$ is the $\chi \chi \varphi$ coupling.

The loop integral in Eq.(\ref{ap1}) can be computed straightforwardly
using Feynman parameters, giving
\begin{equation} \label{ap2}
\frac{A_{\rm vertex}} {A_0} = - \frac {\kappa^2} {16
  \pi^2} C_0(s, m_\chi^2, m_\chi, m_\chi, \mu^2)\,.
\end{equation}
Here $C_0$ is the scalar Passarino--Veltman three--point function in
the convention of ref.\cite{Hollik}.

The loop integral in Eq.(\ref{ap1}) can also be evaluated directly,
following the steps of Sec.~2 but without making any approximations in
the propagators. We first perform the energy ($q_0$) integrals by
contour integration, by summing over the residues of all poles in the
lower half plane. In general, there are three such poles:
\begin{eqnarray} \label{ap3}
q_0^{\rm pole \ 1} &=& \omega - P_0 \, ; \nonumber \\
q_0^{\rm pole \ 2} &=& \omega + P_0 \, ; \nonumber \\
q_0^{\rm pole \ 3} &=& \sqrt{(\vec{p} - \vec{q})^2 + \mu^2}\, ,
\end{eqnarray}
where $\omega = \sqrt{\vec{q}^2 + m_\chi^2}$ as in Eq.(\ref{en2}).
Only the first pole has a residue that diverges in the limit $\vec{p},
\vec{q} \rightarrow 0$. The third pole comes from the energy
dependence of the $\varphi$ propagator, which has been ignored in the
approximate treatment of Sec.~2. The angular integrals can also be
performed straightforwardly. After some algebra, we arrive at:
\begin{eqnarray} \label{ap4}
\frac{A_{\rm vertex}} {A_0} &=& \frac {\kappa^2} {16
  \pi^2} \frac{1}{4 P_0 |\vec{p}|} \int_0^\infty |\vec{q}| d |\vec{q}| 
\left[ \frac  {1}{ \omega (\omega - P_0) } 
\ln \frac { \left( |\vec{p}| + |\vec{q}| \right)^2 + \mu^2 -
  \left(\omega - P_0 \right)^2 }
{ \left( |\vec{p}| - |\vec{q}| \right)^2 + \mu^2 -
  \left(\omega - P_0 \right)^2 } 
\right. \nonumber \\ && \hspace*{37mm} \left. 
-\ \frac  {1}{ \omega (\omega + P_0) }
\ln \frac { \left( |\vec{p}| + |\vec{q}| \right)^2 + \mu^2 -
  \left(\omega + P_0 \right)^2 }
{ \left( |\vec{p}| - |\vec{q}| \right)^2 + \mu^2 -
  \left(\omega + P_0 \right)^2 } 
\right. \\ && \hspace*{37mm}\left. 
+\  \frac{1}{\omega_\varphi^2} \left( \ln \frac {\mu^2 + 2 |\vec{p}|
    |\vec{q}| - 2 P_0 \omega_\varphi} {\mu^2 - 2 |\vec{p}|
    |\vec{q}| - 2 P_0 \omega_\varphi}
- \ln \frac {\mu^2 + 2 |\vec{p}|
    |\vec{q}| + 2 P_0 \omega_\varphi} {\mu^2 - 2 |\vec{p}|
    |\vec{q}| + 2 P_0 \omega_\varphi} \right) \right] \,. \nonumber
\end{eqnarray}
In the last line, we have introduced $\omega_\varphi = \sqrt{\vec{q}^2
  + \mu^2}$.

It is easy to see that the first term reduces to our expression
(\ref{eq:deltaAS}) if we use the non--relativistic expansion for
$\omega$, which includes dropping the terms $(\omega - P_0)^2$ in the
argument of the logarithm. However, for large $|\vec{q}|$ these latter
terms are important. They imply that the logarithm approaches the
constant value $2 |\vec{p}|/P_0$ in the limit $|\vec{q}| \rightarrow
\infty$, rather than vanishing as in Eq.(\ref{eq:deltaAS}). As a
result, the first line of the right--hand side (rhs) of Eq.(\ref{ap4})
by itself is logarithmically UV--divergent. The second line
contributes the same UV divergence again; only after adding the
contribution in the third line we obtain a UV--finite result. This
third line comes from the third pole in Eq.(\ref{ap3}), which does not
exist if one drops the energy dependence of the $\varphi$ propagator,
as in Sec.~2. This proves our statement in Sec.~2 that omission of the
energy dependence of the $\varphi$ propagator necessitates the use of
a non--relativistic expansion in the argument of the loop integral.

While the third line in Eq.(\ref{ap4}) is necessary to obtain a
UV--finite result, it introduces a new problem: for $\mu \rightarrow
0$ it becomes IR--divergent! In the limit $\mu^2 \ll m_\chi^2, \
|\vec{p}| \ll P_0$ the last line of Eq.(\ref{ap4}) simplifies to
$$ \frac {16 P_0 |\vec{p}| |\vec{q}| } {\sqrt{\vec{q}^2 + \mu^2} } 
\frac{1}{ \mu^4 - 4 P_0^2 \left( \vec{q}^2 + \mu^2 \right) }\,.$$ The
$d |\vec{q}|$ integration will then lead to a {\em negative} term
$\propto \ln \frac {|\vec{q}|_{\rm max}}{\mu}$.  As noted earlier, the
UV divergence for $|\vec{q}|_{\rm max} \rightarrow \infty$ precisely
cancels those from the first two terms in Eq.(\ref{ap4}). The
resulting term $\propto \ln \frac {m_\chi}{\mu}$ in the exact vertex
correction is IR--divergent for $\mu \rightarrow 0$. This term becomes
significant for small $\mu$, especially if the velocity $v$ is not too
small. This explains our statement in Sec.~2 that our approximation
does {\em not} describe the exact vertex correction very well for
small $\mu$.

However, the IR divergence does not exist in the {\em full}
one--loop calculation. We have to add wave function renormalization
(Fig.~8b) as well as real emission diagrams (Fig.~8c) in order to
obtain an IR--finite result for $\mu \rightarrow 0$. This implies that
adding these additional contributions should also remove all terms
$\propto \ln \frac {m_\chi} {\mu}$ in the complete one--loop corrected
cross section. Using on--shell renormalization for $\chi$, the wave
function renormalization constant $Z_\chi$ is finite:\footnote{Diagram
  8b also gives a logarithmically divergent contribution to
  $m_\chi$. This is simply removed by the mass counterterm in
  on--shell renormalization.}
\begin{equation} \label{ap5}
Z_\chi = \frac{\kappa^2}{16 \pi^2} B_0'(m_\chi^2,m_\chi, \mu)\,,
\end{equation}
where $B_0'$ is the derivative of the scalar Passarino--Veltman
two--point function with respect to its first argument. Adding this
negative contribution {\em doubles} the IR--divergence for $\mu
\rightarrow 0$.

Finally, we have to treat the real $\varphi$ emission diagram of
Fig.~8c, plus the contribution where $\varphi$ is emitted off the
other $\chi$ line. Writing the 4--momentum of the emitted $\varphi$ as
$k = (k_0, \vec{k})$, we have
\begin{equation} \label{ap6}
\frac { \left| A_{\rm real \ em.} \right|^2} { \left| A_0 \right|^2 }
= \kappa^2 \left| \frac { 2 \left( \mu^2 - 2 k_0 P_0 \right) } {
    \left( \mu^2 - 2 k_0 P_0 \right)^2 - 4 \left( \vec{k} \cdot
      \vec{p} \right)^2 } \right|^2\,. 
\end{equation}
Performing the angular integrations of the $\varphi$ phase space, this
gives
\begin{eqnarray} \label{ap7}
\frac {\sigma_{\rm real \ em.}} {\sigma_0} &=& \frac {\kappa^2}{2 \pi^2}
\int_\mu^{k_{0,{\rm max}}} d k_0   \left[ \frac { |\vec{k}|} 
{ \left( 2 P_0 k_0 - \mu^2 \right)^2 - 4 \vec{k}^2 \vec{p}^2}
\right. \nonumber \\ && \hspace*{26mm}\left.
+\ \frac {1} {4 |\vec{p}| \left( 2 P_0 k_0 - \mu^2 \right) }
\ln \frac { 2 k_0 P_0 - \mu^2 + 2 |\vec{k}| |\vec{p}| }
 { 2 k_0 P_0 - \mu^2 - 2 |\vec{k}| |\vec{p}| } \right]\,,
\end{eqnarray}
where $|\vec{k}| = \sqrt{k_0^2 - \mu^2}$ and $k_{0,{\rm max}} = P_0 +
\mu^2/(4 P_0)$. In the limit $\mu \rightarrow 0$ this also produces a
logarithmic IR divergence, this time with positive sign, which (not
surprisingly) cancels the sum of the IR divergent terms from the
vertex correction and wave function renormalization.

\begin{figure}[h!]
\centerline{\rotatebox{270}{\includegraphics[width=130mm]{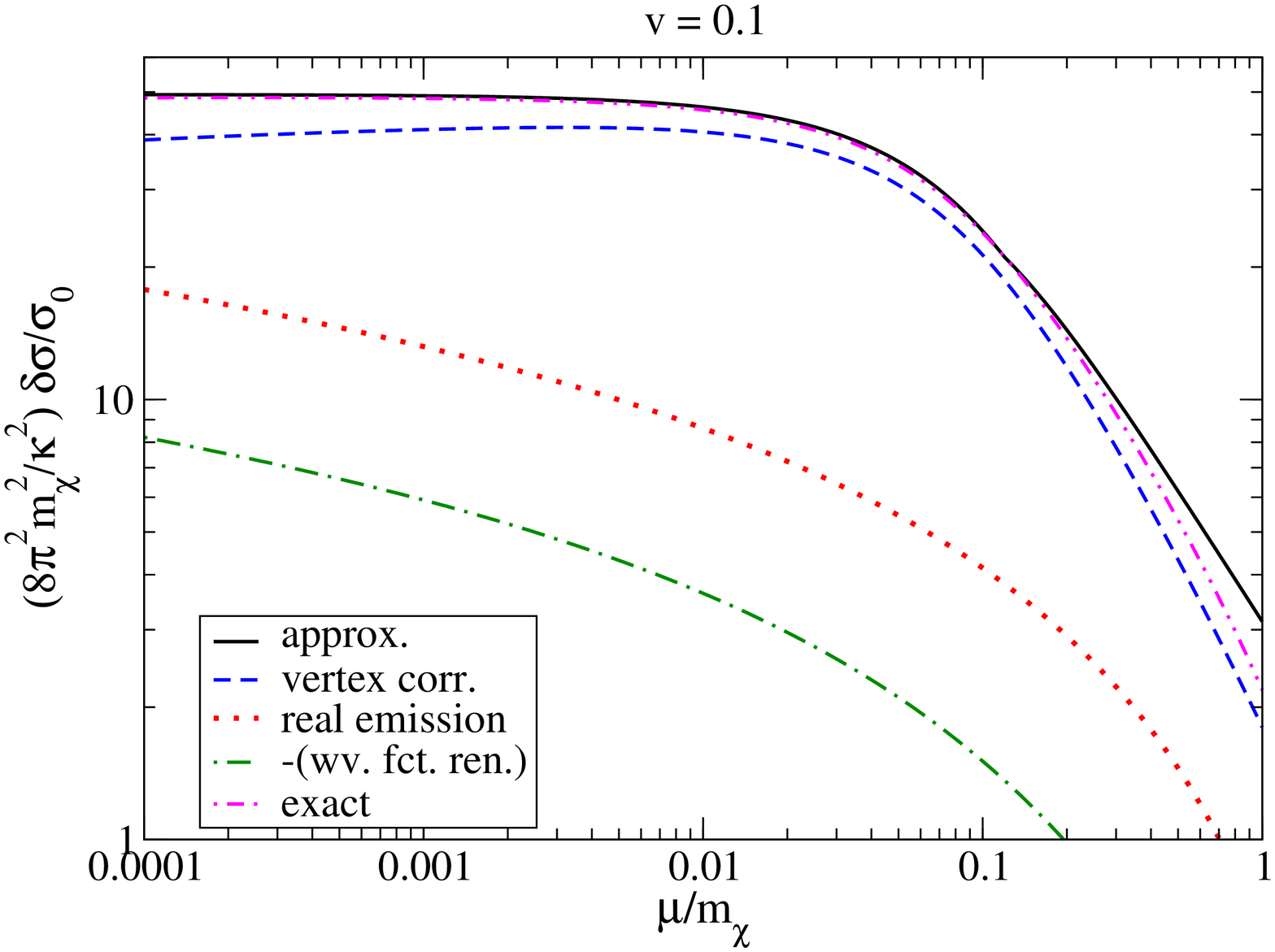}}}
\vspace*{1mm}
\caption{Normalized higher order contributions to the $\chi$
  annihilation cross section for relative initial state velocity $v =
  0.1$. The solid (black) curve shows our approximation of Sec.~2,
  given by $I_S/v$. The dashed (blue), dotted (red) and dot--dashed
  (green) curves show exact contributions from vertex corrections,
  real emission diagrams, and wave function renormalization,
  respectively; the latter has to be multiplied with $-1$. The
  dash--doubledotted (magenta) curve shows the sum of these three
  contribution, i.e. the exact one--loop correction; it nearly
  coincides with the black curve for $\mu / m_\chi \leq 0.2$.}
\label{fig:comp1}
\end{figure}

Fig.~\ref{fig:comp1} shows that the sum of the vertex correction, wave
function renormalization and real emission contributions very closely
matches our approximate result of Sec.~2 for $\mu \lsim 0.5
m_\chi$. In fact, the difference is always of order $\kappa^2 / (8
\pi^2 m_\chi^2)$, without any potentially large factors like $1/v$ or
$\ln(m_\chi/\mu)$. We checked that this remains true at least for all
$v \lsim 0.5$ relevant for the calculation of the relic
density.\footnote{Our approximation might fail badly when the
  annihilating WIMPs become ultra--relativistic, but this is of no
  concern in the present context.} It is not surprising that our
approximate treatment does not treat such ``generic'' higher order
contributions correctly. However, our approximation does closely
resemble the exact result whenever the latter is large. This is all we
aspired to, and in most cases all we need when calculating DM relic
densities, even if PLANCK data reduce the uncertainty of the observed
value to the percent level.

\end{document}